\documentclass[a4paper,11pt]{article}
\pdfoutput=1 
\usepackage{jheppub} 
\usepackage{graphicx,color}
\usepackage[dvipsnames]{xcolor}
\usepackage{amsmath}
\usepackage{autobreak}
\allowdisplaybreaks
\usepackage{floatrow}
\DeclareUnicodeCharacter{2212}{-}

\DeclareFloatFont{tiny}{\tiny}
\def\bt#1{\beta_{#1}}

\def\z#1{\zeta_{#1}}
\def\D#1{{{{\color{MidnightBlue}\boldsymbol{\cal D}_{#1}}}}}
\def\Dm1{{{\delta(1-z)}}}
\def\Dm1{{{{\color{MidnightBlue}\boldsymbol{\delta(1-z)}}}}}

\def\CIg#1{{{\Delta_{gg,sv}^{(#1),GR}}}}
\def\CIq#1{{{\Delta_{q\bar{q},sv}^{(#1),GR}}}}
\def\LogmW1{{{\ln (1-\omega)}}}

\def\g0#1DY{{g_{0#1}^{DY}}}
\def\gGRGG#1{{g_{0,g}^{(#1),GR}}}
\def\gGRQQ#1{{g_{0,q}^{(#1),GR}}}

\newcommand{\nn}{\nonumber}
\newcommand{\eps}{\epsilon}
\newcommand{\Ca}{C_A}
\newcommand{\Cf}{C_F}
\newcommand{\nf}{n_f}

\newcommand{\Lqf}{L_{qf}}
\newcommand{\olsi}[1]{\,\overline{\!{#1}}} 
\newcommand{\Nb}{{\olsi{N}}}

\newcommand{\rqg}{r_{g}}
\newcommand{\rgq}{r_{q}}
\newcommand{\grg}{r_g}
\newcommand{\grq}{r_q}
\newcommand{\omg}{\omega}
\newcommand{\Lqrfrt}{L_{qr}L_{fr}^{2}}
\newcommand{\Lqrfr}{L_{qr}L_{fr}}
\newcommand{\Lqrtfr}{L_{qr}^{2}L_{fr}}

\newcommand{\Lqr}{L_{qr}}
\newcommand{\Lfr}{L_{fr}}

\newcommand{\psih}{\hat{\psi}}
\newcommand{\psihbar}{\overline{\psih}}
\newcommand{\Ah}{\hat{A}}
\newcommand{\omgh}{\hat{\omega}}
\newcommand{\omghbar}{\overline{\omgh}}
\newcommand{\xih}{\hat{\xi}}
\newcommand{\fabc}{f^{abc}}
\newcommand{\gmunu}{g_{\mu\nu}}
\newcommand{\Fh}{\hat{F}}
\newcommand{\gh}{\hat{g}_{s}}
\newcommand{\gmu}{\gamma_\mu}

\newcommand{\gsig}{\gamma^{\sigma}}
\newcommand{\del}{\partial}

\newcommand{\eq}[1]{eq.\ (\ref{#1})}
\newcommand{\fig}[1]{fig.\ (\ref{#1})}

\newcommand{\sect}[1]{sec.\ (\ref{#1})}
\newcommand{\app}[1]{appendix\ \ref{#1}}

\begin{document} 
\title{Precision QCD phenomenology of exotic spin-2 search at the LHC}
\author[a]{Goutam Das,}
\author[b]{M. C. Kumar }
\author[b]{ and Kajal Samanta }
\affiliation[a]{Theoretische Physik 1, 
Naturwissenschaftlich-Technische Fakult{\"a}t, 
Universit{\"a}t Siegen, Walter-Flex-Strasse 3, 57068 Siegen,
 Germany}
\affiliation[b]{Department of Physics, Indian Institute of 
Technology Guwahati, Guwahati-781039, Assam, India}
\emailAdd{goutam.das@uni-siegen.de}
\emailAdd{mckumar@iitg.ac.in}
\emailAdd{kajal.samanta@iitg.ac.in }

\date{\today}

\abstract{
 The complete next-to-next-to leading order (NNLO) QCD 
 correction matched with next-to-next-to leading logarithm 
 (NNLL)  has been studied for Drell-Yan production through 
  spin-2 particle at the Large hadron collider 
 (LHC). We consider generic spin-2 particle which couples 
 differently to the quarks and the gluons (non-universal 
 scenario). The threshold enhanced analytical coefficient has 
 been obtained up to third order exploiting the universality of
 the soft function as well as the process dependent form 
 factors at the same order. We performed a detailed 
 phenomenological analysis and give prediction for the 13 
 TeV LHC for the search of such BSM signature. We found that 
 the matched correction at the second order gives sizeable 
 corrections over wide range of invariant mass of 
 the lepton pair.
 The scale variation also stabilizes at 
 this order and reduces to $4\%$. As a by-product we 
 also provide ingredients for third order soft-virtual (SV) prediction as 
 well as resummation and study the impact on LHC searches.
 }

\preprint{SI-HEP-2020-32, P3H-20-077}
\maketitle
\section{Introduction}\label{sec:introduction}
Following the discovery of the Higgs boson at the 
Large Hadron Collider (LHC) 
\cite{Aad:2012tfa,Chatrchyan:2012ufa}, 
much of the focus is given to the search for signals of possible 
beyond Standard Model (BSM) Physics scenarios. 
There have been many such models 
proposed to address various problems in the Standard Model (SM) 
{\it viz.} 
supersymmetry, extra dimensions, little Higgs, technicolor etc. 
\cite{Zyla:2020zbs}. These 
signals can be seen either in the form of smooth deviations from the 
SM predictions due to contact interactions, or in the form of 
resonances due to the existence of new heavy particles that these 
models predict.

In the context of LHC, there have been various 
observable like invariant mass and transverse momentum of the final states
involving either leptons or photons that have a very large detection
efficiency ($\ge$ 90\%) at the ATLAS and CMS detectors.  
Among such observable, the dilepton invariant mass is of 
particular interest as the lepton (electron and muon) signals are very clean
and the invariant mass ($Q$) 
of the lepton pair has been measured up to 2 TeV to a very good accuracy
\cite{Aaboud:2017buh,Sirunyan:2018ipj,Khachatryan:2016zqb}. 
It is
to be noted that with the availability of such increasing precision in the
experimental data, an adequate theoretical prediction to that accuracy 
is necessary. After the computation of the Higgs production to 
N$^3$LO accuracy \cite{Anastasiou:2015vya},
such a high precision has been achieved for the dilepton production 
at hadron
colliders, thanks to
the recent computation  in the photon mediated
 channel \cite{Duhr:2020seh}
and the massive gauge boson $W^\pm$ production \cite{Duhr:2020sdp}.  
Such a precision 
both in the theoretical as well as in the experimental frontiers 
facilitates
a robust search for BSM signals.

In this work, we focus on one particular BSM scenario where generic massive 
spin-2 fields interact with the SM ones. Such a model is well motivated in the 
context of search for spin-2 particles (graviton) that couple to 
SM bosons and fermions 
differently. 
The phenomenology of such a graviton is similar to some extent 
to that
of the RS model \cite{Randall:1999ee} where the massive graviton 
couples with equal 
strength to fermions
and bosons in the SM. However unlike the RS model,
the parameter space of the non-universal case is much flexible 
and less constrained at the LHC searches so far.
In the context of 
Higgs characterisation,
this model has been studied extensively in the di-boson channels 
\cite{Artoisenet:2013puc}.
Later, a complete automation has been done at NLO \cite{Das:2016pbk}
in the 
F{\sc eyn}R{\sc ules} \cite{Alloul:2013bka}-M{\sc ad}G{\sc raph}5\_{\sc a}MC@NLO \cite{Alwall:2014hca} 
framework.
A detailed phenomenology has been performed there   
for arbitrary values of mass of graviton and its couplings to 
the SM including parton shower effects as well.
It has been observed that the K-factors for different channels 
give sizeable 
contribution depending on model parameters as well as 
phase space region. This necessitates further
higher order QCD corrections for this model.  
The first complete next-to-next-to leading order (NNLO) computation 
has been performed  \cite{Banerjee:2017ewt} in the
dilepton channel. A detailed analysis has been presented there
along with the comparison with universal scenarios.
It has been observed that the K-factors
at NLO and NNLO are significantly different from those of
 the ADD \cite{ArkaniHamed:1998rs}  or RS \cite{Randall:1999ee} cases.
Note that both in ADD and RS models, the K-factors for the dilepton production are found to be 
much larger than those in the SM.  This behaviour is
 well understood because of the
presence of the additional gluon fusion process 
(Higgs-like) at the born level in addition to the quark 
annihilation process (Drell-Yan-like) for graviton production, which is absent
in the SM where the vector boson production takes place via only quark annihilation process
at the LO. 
In the present model, the graviton production takes
 place at the LO via Higgs-like and
DY-like processes similar to the ADD or RS model, but with different couplings to 
quarks ($k_q$) and gluons ($k_g$). 
This particular feature of this model controls the 
size of the contributions from different channels 
at higher orders in different parameter space.

The observed discrepancy between the K-factors for the graviton production in
the universal and the non-universal cases can be explained to some extent with the help
of additional parton radiations at higher orders. For example, a DY-like process with 
small $k_q$ coupling can receive large corrections when the graviton couples
with large coupling $k_g$ to gluon emitted from the quark lines at higher orders.
Similar is the case for Higgs-like graviton process. The contribution of other parton level
subprocesses initiated by $qg$ starting from NLO and 
$qq\prime$ from NNLO onwards will have
further noticeable contributions. 
For example at NLO, the $q\bar{q}$ subprocess gets  corrections a 
few times larger than the LO subprocess  for the choice of parameters 
where $k_q \ll k_g$ simply due to the fact that high $k_g$ contribution
 appear at NLO from $q\bar{q}$ subprocess itself.  
The contribution from the $qg$ channel could be as large as $q\bar{q}$
for different parameter choices.
The presence of such model dependent
large contributions at higher orders for an inclusive process can not be estimated with a simple scaling of the lower
order cross sections with the conventional (NLO/NNLO) K-factors computed either in the SM or 
in the RS model. This motivates not only a detailed phenomenological study at higher orders in QCD
but also questions the convergence of the perturbation theory. 
This is particularly important in the higher mass region 
where one would 
normally expect the BSM effect. This region
suffers from large threshold corrections and needs a 
procedure to resum them to all orders. 
The resummation procedure has been pioneered 
\cite{Sterman:1986aj,Catani:1990rp,
Catani:2003zt,Eynck:2003fn,Moch:2005ba,
Becher:2006mr,Catani:2014uta,Ahmed:2020nci} over the
last few decades to systematically include those large
 contributions
in the threshold region from all orders providing
better predictability of the perturbative series. 


At the LHC energies where the parton fluxes are large,
the threshold logarithms can give sizeable contribution 
and through resummation these lead to better perturbative convergence as well as 
theoretical uncertainties for various processes 
\cite{Catani:2003zt,Moch:2005ky,Moch:2005ba,deFlorian:2007sr,
Beenakker:2009ha,Kidonakis:2010tc,Bauer:2011uc,
Harlander:2014wda,Catani:2014uta,Bonvini:2014joa,Bonvini:2015pxa,Ahmed:2015sna,
Banfi:2015pju,Schmidt:2015cea,Ahmed:2016otz,Bonvini:2016frm,H:2019dcl,
Kramer:2019wvd,Das:2019btv,H.:2020ecd,Das:2020adl}. 
Recently the effect of these large threshold logarithms has been studied in ADD and RS models 
\cite{Das:2019bxi,Das:2020gie} and better perturbative results have been predicted 
at N$^3$LL(NNLL) accuracies.
In view of this, we systematically include the threshold
 resummation effects in DY production
 for non-universal spin-2 model and present
a detailed phenomenological results 
up to NNLL accuracy.
Moreover we have also computed the necessary ingredients to perform 
resummation up to N$^3$LL.

The paper is organized as follows. 
In the \sect{sec:theory}, 
we develop the theoretical formalism and collect 
important formulas needed for DY production at the LHC. 
In \sect{sec:numerics},
we present detailed numerical results for the resummed 
cross-section matched to the fixed order at the NNLO level.  
We also present a quantitative estimate of the threshold
corrections at the fixed order N$^3$LO level and 
finally give a summary of findings in \sect{sec:conclusion}.

\section{Theoretical Framework}\label{sec:theory}
The  interaction of spin-2 field ($h_{\mu\nu}$) with the SM ones is 
described through the following effective action, 
\begin{align}\label{eq:interaction-action}
S_{int} = -\frac{1}{2}\int d^4x~ h_{\mu\nu}(x) \left( 
\hat{k}_I \hat{\mathcal{O}}^{I,\mu\nu}(x) 
\right )\,, ~\qquad I\in\{G,Q\}\,.
\end{align}
Sum over repeated indices is implied in all equations. The interaction 
action is splitted in a way so that $I=G$ contains purely gauge terms 
whereas $I=Q$ contains the fermionic sector and its gauge interactions 
terms. This decomposition however is not unique and one can shuffle gauge 
invariant terms between these two. $\hat{k}_I$ are the unrenormalized 
coupling constants with which the spin-2 field  couples to the operators 
$\hat{\mathcal{O}}^{I,\mu\nu}$. These gauge invariant operators have the 
following expressions in terms of unrenormalized quarks and gluon 
fields\footnote{Throughout this article we use \textit{hat} ($~ \hat{} ~$) 
symbol for bare quantities and without \textit{hat} as renormalized 
quantities.},
\begin{align}\label{eq:unrenorm-operators}
\begin{autobreak}
 \hat{{\cal O}}^Q_{\mu\nu} =  
    \frac{i}{4} \psihbar \Big\{ \gmu\overrightarrow{D_\nu}  -  \overleftarrow{D_\nu}\gmu \Big\}  \psih
 -  \frac{i}{2} \gmunu \psih \gsig \overrightarrow{D_\sigma} \psih  
 +~ \Big(\mu \leftrightarrow \nu\Big)\,,
\end{autobreak}
\\
\begin{autobreak}
 \hat{{\cal O}}^G_{\mu\nu} =
    \frac{1}{4} \gmunu\Fh_{\rho\sigma}^a\Fh^{a,\rho\sigma}  
 -  \Fh_{\mu\rho}^a \Fh_\nu^{a,\rho}
 -  \frac{1}{\xih} \gmunu \del^\rho (\Ah_\rho^a \del^\sigma \Ah_\sigma^a)
 -  \frac{1}{2\xih} \gmunu (\del_\rho \Ah^{a,\rho}) (\del_\sigma \Ah^{a,\sigma})
 +  \bigg\{\frac{1}{\xih} \Ah_\mu^a \del_\nu(\del^\rho \Ah_\rho^a) 
 +  \del_\mu \omghbar (\del_\nu\omgh^a - \gh\fabc\Ah_\nu^c\omgh^b) 
 -  \frac{1}{2}\gmunu \del_\rho \omghbar^a (\del^\rho \omgh^a - \gh\fabc\Ah_{c,\rho}\omgh^b) 
 + \mu \leftrightarrow \nu\bigg\}\,,
\end{autobreak}
\end{align}
where $\overrightarrow{D_\mu} = \overrightarrow{\del_\mu} - i \gh T^a \Ah_\mu^a$. $\omgh$ and $\xih$ are the ghost field and gauge fixing parameter respectively. Notice that the sum of these operators are protected against radiative corrections to all orders due to the fact that it corresponds to the conserved energy momentum tensor of QCD. However individually they are not conserved and requires additional UV renormalization. These operators are closed under UV renormalization and  hence renormalization can be performed \cite{Joglekar:1975nu,Henneaux:1993jn} through a mixing matrix $Z$.

The unrenormalized operators $\hat{\mathcal{O}}_I^{\mu\nu}(x)$ can be 
renormalized in a closed form \cite{Joglekar:1975nu,Henneaux:1993jn} using 
renormalization matrix $Z_{IJ}$ as
\begin{align}\label{eq:operator-renorm}
 \mathcal{O}_I  = Z_{IJ}~ \hat{\mathcal{O}}_J\,.
\end{align}
We use dimensional regularization in $d=4-2\eps$ dimensions to regulate 
both UV and IR divergences. In $d$-dimensions, the bare strong coupling 
can be related to the renormalized one as 
\begin{align}\label{eq:as-renorm}
 \hat{a}_s S_d = a_s(\mu_r) Z(a_s(\mu_r)) \left(\frac{\mu_r^2}{\mu^2} \right)^{\frac{4-d}{2}}\,,
\end{align}
where $S_d$ is the spherical factor. $Z(a_s(\mu_r))$ is the strong 
coupling renormalization constant which takes the following form in 
$d=4-2\eps$ dimensions.
\begin{align}\label{eq:as-renorm-constant}
Z(a_s(\mu_r)) = 1 +& a_s(\mu_r)\Big\{
-\bt0 \eps^{-1}
\Big\} + 
a_s^2(\mu_r)\Big\{
\bt0^2\eps^{-2} 
-
\frac{\bt1}{2}\eps^{-1}
\Big\} + 
a_s^3(\mu_r)\Big\{
-\bt0^3 \eps^{-3}
\nonumber\\
+& \frac{7}{6}\bt0\bt1 \eps^{-2}
- \frac{1}{3} \bt2 \eps^{-1}
\Big\}\,.
\end{align}
The UV renormalization constants $Z_{IJ}$ satisfy the following RGE
\begin{align}\label{eq:uv-renorm-rge}
 \frac{d}{d \ln \mu_r} Z_{IJ} = \gamma_{IK} Z_{KJ}\,,
\end{align}
where $\gamma_{IJ}$ are the UV anomalous dimensions.

The interation action can be expressed in terms of renormalized quatities
as well \textit{i.e.} in terms of renormalized couplings ($\kappa_I$) and 
renormalized operators ($\mathcal{O}_I$)\footnote{Note that in a pure gauge 
theory ($n_f=0$), the operator $\mathcal{O}_G^{\mu\nu}$ is conserved.} as
\begin{align}\label{eq:renorm-action}
S_{int} = -\frac{1}{2}\int d^4x~ h_{\mu\nu}(x) \Big( 
\kappa_I \mathcal{O}_I^{\mu\nu}(x) 
\Big) \,.
\end{align}
The unrenormalized coupling constants are then related to the renormalized 
ones with the transpose of the same renormalization constants\footnote{We 
confirmed a minor typo in eq.\ (2.11) of \cite{Ahmed:2016qjf} which however 
finally does not affect the results there.} $Z_{IJ}$  as
\begin{align}\label{eq:coupling-renorm}
\hat{\kappa}_I = Z_{JI}~ \kappa_J \,.
\end{align}
The solution of  UV renormalization constants RGE (\eq{eq:uv-renorm-rge}) 
matrix leads to the following expansion in terms of renormalized strong 
coupling up to the third order,
\begin{align}\label{eq:uv-renorm-const}
Z_{IJ} = \delta_{IJ} +
a_s(\mu_r) \Big\{ 
\Big(
&-\gamma_{IJ}^{(1)}
\Big) \eps^{-1} \Big\}
+ a_s(\mu_r)^2 \Big\{
\Big( \frac{1}{2}\bt0 \gamma_{IJ}^{(1)} + \frac{1}{2} \gamma_{IK}^{(1)} \gamma_{KJ}^{(1)} 
\Big) \eps^{-2} +  \Big( -\frac{1}{2}\gamma_{IJ}^{(2)}
\Big) \eps^{-1} \Big\} 
\nonumber\\
+ a_s(\mu_r)^3 \Big\{
\Big(
&- \frac{1}{3}\bt0^2 \gamma_{IJ}^{(1)}
- \frac{1}{2} \bt0 \gamma_{IK}^{(1)}\gamma_{KJ}^{(1)} -
    \frac{1}{6}\gamma_{IK}^{(1)}\gamma_{KL}^{(1)}\gamma_{LJ}^{(1)}
\Big)\eps^{-3} 
+
\Big( \frac{1}{3} \bt1 \gamma_{IJ}^{(1)} 
    \frac{1}{3} \bt0 \gamma_{IJ}^{(2)} 
\nonumber\\    
 &+   \frac{1}{6} \gamma_{IK}^{(1)} \gamma_{KJ}^{(2)} +
    \frac{1}{3} \gamma_{IK}^{(2)} \gamma_{KJ}^{(1)}
\Big)\eps^{-2} 
+\Big(
-\frac{1}{3} \gamma_{IJ}^{(3)}
\Big)\eps^{-1}
\Big\} \,.
\end{align}
 All the relevant anomalous dimensions ($\gamma_{IJ}$) are extracted to 
 three loops \cite{Banerjee:2017ewt} from the bare quark and gluon form 
 factors and by simply claiming the universality of the infrared 
 divergences at the same order. For the sake of completeness we have 
 collected those in the appendix \app{app:anomalous-dimensions}.
Note that one can equivalently perform coupling constant renormalization 
using \eq{eq:coupling-renorm} instead of operator renormalization in 
\eq{eq:operator-renorm} to remove the UV divergences from the form factor.

We now turn our discussion into the DY production cross-section at the LHC 
which takes the following form in terms of partonic coefficient function 
$\Delta^I$ and luminosity $\mathcal{L}$,
\begin{align}\label{eq:hadronic-xsect}
2 S{d \sigma \over d Q^2}\left(\tau,Q^2\right)
=\sum_{ab=\{q,\overline q,g\}} \int_0^1 dx_1
&\int_0^1 dx_2
\int_0^1 dz \,\, 
\mathcal{L}_{ab}(x_1,x_2,\mu_f^2) \times
\nonumber\\ 
&\sum_I \mathcal{F}_I^{(0)} \Delta_{a b}^{I}(z,Q^2,\mu_f^2)
\delta(\tau-z x_1 x_2)\,.
\end{align}
 The luminosity function consists of non-perturbative parton distribution 
 functions. The prefactor $\mathcal{F}_I^{(0)}$ is given for SM and for 
 spin-2 (denoted as GR) channels as follows
 \begin{align}\label{eq:partonic-prefactor}
\begin{autobreak}
\mathcal{F}_{SM}^{(0)} 
= \frac{4 \alpha^2}{3 Q^2} \Bigg[Q_q^2 
- \frac{2 Q^2 (Q^2-M_Z^2)}{\Big((Q^2-M_Z^2)^2 + M_Z^2 \Gamma_Z^2\Big) c_w^2 s_w^2} Q_q g_e^V g_q^V 
+ \frac{Q^4}{\Big((Q^2 - M_Z^2)^2 + M_Z^2 \Gamma_Z^2\Big) c_w^4 s_w^4}\Big((g_e^V)^2 + (g_e^A)^2\Big)\Big((g_q^V)^2+(g_q^A)^2\Big) \Bigg]\,.
 \end{autobreak}
 \\
  \begin{autobreak}
\mathcal{F}_{GR}^{(0)}  
= \frac{k_q^2 Q^6}{80\pi^2 \Lambda^4} |D(Q^2)|^2\,,
\end{autobreak}
 \end{align}
 Here $\alpha$ represents the fine structure constant, $c_w,s_w$ are the 
 sine and cosine of the Weinberg angle and $M_Z$ and $\Gamma_Z$ are the 
 mass and width of the $Z$ boson respectively. The vector and axial 
 coupling of the weak boson is given as
\begin{align}
g_a^A = -\frac{1}{2} T_a^3 \,, \qquad g_a^V = \frac{1}{2} T_a^3  - s_w^2 Q_a \,,
\end{align}
 $\Lambda_{GR}$ in \eq{eq:partonic-prefactor} is the cut-off scale 
 of the spin-2 theory and $k_I$ are introduced for convenience and are 
 defined as $k_I = \sqrt{2}\kappa_I/\Lambda$. The propagator for GR 
 theory with mass $M_G$ is given as
 \begin{align}\label{eq:dc-propagator}
  D(Q^2) = \frac{1}{(Q^2 - M_G^2) + i ~ \Gamma_G M_G} \,.
 \end{align}

 The analytic form of the partonic coefficient function is known for some 
 time to the second order for all subprocesses. The partonic coefficients 
 can be decomposed into the following form for the gluon and 
 quark-antiquark initiated processes:
\begin{align}\label{eq:partonic-decomposition}
 \Delta_{ab}^{I} = \Delta_{ab,sv}^I + \Delta_{ab,reg}^I \,, ~ \qquad  ab \in \{ g g, q\bar{q} \} \,. 
\end{align}
The first term in the above equation is termed as the soft-virtual term 
which consists of form factor contribution as well as soft gluon radiation.
The perturbative expansion of the SV coefficients are given as,
\begin{align}\label{eq:sv-expansion}
 \Delta_{ab,sv}^{I} = \sum_{i=0} a_s^i \mathcal{N}_{ab}^I \Delta^{(i),I}_{ab,sv}
\end{align}
$\mathcal{N}_I$ are some overall prefactors taken out from the SV coefficient. The $\mathcal{N}_I$ 
prefactors are given as
\begin{align}\label{eq:prefactor-extra}
\mathcal{N}_{ab}^{SM}       &= \frac{2\pi}{N_c}\,, \nonumber\\
\mathcal{N}_{q\bar{q}}^{GR} &= \frac{\pi k_q^2}{8 N_c}\,,  \nonumber\\
\mathcal{N}_{gg}^{GR}       &= \frac{\pi k_g^2}{2(N_c^2-1)}\,.
\end{align}
Note that with this normalisation in \eq{eq:prefactor-extra}, the SV 
coefficient at LO simply becomes $\delta(1-z)$. The main ingredients to 
compute the SV coefficients (after the PDF renormalization) are the form 
factor and universal soft distribution functions. As mentioned earlier, 
the form factors are available in the literature up to three loops 
\cite{Ahmed:2016qjf}.
We have used the \eq{eq:coupling-renorm} to remove the UV divergences 
from the bare form factors. The UV finite form factor then contains the 
IR divergences originating from the soft-collinear region. After performing 
collinear renormalization of parton distribution functions, the remaining 
IR divergences are cancelled upon inclusion of universal soft distributions. The universal soft distributions are already known in the literature up to three loops \cite{Anastasiou:2014vaa,Catani:2014uta,Laenen:2005uz,Ahmed:2014cla,Li:2014bfa}. This finally leads to finite SV coefficients up to third order which contains $\delta(1-z)$ and plus-distributions $\mathcal{D}_n = \Big[\frac{\ln^{2n-1}(1-z)}{1-z}\Big]_+$ in partonic threshold variable $z=Q^2/\hat{s}$. The new third order SV coefficients are collected in the \app{app:sv-coefficints}.

In order to better describe the fixed order cross-section particularly in 
the threshold region, one needs to resum threshold enhanced logarithms to 
all orders. In the threshold region the partonic $z\to1$ induces large 
singular terms from delta function and plus distributions. The resummation 
is performed in Mellin space where the threshold limit $z\to1$ translates 
into $N\to\infty$ and the large logarithms appear in the form of 
$\ln^i \Nb$ (with $\Nb = N\exp(\gamma_E)$, $\gamma_E$ is Euler-Mascheroni 
constant.). The partonic coefficient takes the following exponential form 
(up to normalisation by born factor.)
\begin{align}\label{eq:resum-partonic}
(d\hat{\sigma}^I_N/dQ)/(d\hat{\sigma}^I_{\rm LO}/dQ) = g_{0,p}^I \exp \Big( G_{\Nb}^I \Big) \,.
\end{align}
The born normalisation factor is given as
\begin{align}
(d\hat{\sigma}^I_{\rm LO}/dQ) &= {\cal F}^{(0)}_{SM} \frac{Q}{S} \bigg\{ \frac{2\pi}{N_c}\bigg\} \, ~ \qquad\qquad\qquad~ \text{for SM,} \nn\\
 &= {\cal F}^{(0)}_{GR} \frac{Q}{S} \bigg\{ \frac{\pi k_q^2}{8 N_c}, \frac{\pi k_g^2}{2(N_c^2-1)}\bigg\} \qquad \text{for} ~~ \{q\bar{q},gg\}~ \text{in GR.}
\end{align}
The prefactor $g_{0,p}^I$ can be expanded in terms of strong coupling as
\begin{align}\label{eq:g0-expansion}
 g_{0,p}^I = 1 + \sum_{i=1} a_s(\mu_r)^i g_{0,p}^{(i),I}\,, ~ \qquad \qquad 
 p \in \{q,g\} \,.
\end{align}

The universal resummed exponent in \eq{eq:resum-partonic} can be found 
from the Mellin-transform of well known universal cusp anomalous 
dimensions $A_p^I$  \cite{Moch:2004pa, Lee:2016ixa, Moch:2017uml, 
Grozin:2018vdn, Henn:2019rmi, Bruser:2019auj, Davies:2016jie, 
Lee:2017mip, Gracey:1994nn, Beneke:1995pq,  Moch:2018wjh, Lee:2019zop, 
Henn:2019swt,Huber:2019fxe,vonManteuffel:2020vjv} and constant 
$D_p^I$ \cite{Catani:2003zt,Moch:2005ba,Das:2019bxi},
\begin{align}\label{eq:form-of-GN} 
G_{\Nb}^I = \int_0^1 dz \frac{z^{N-1} - 1}{1 - z} \bigg[ \int_{\mu_f^2}^{Q^2(1-z)^2} \frac{d\mu^2}{\mu^2} 2~A_p^I(a_s(\mu^2)) + D_p^I(a_s(Q^2(1-z)^2))\bigg] \,.
\end{align}
This integration can be performed after expanding the anomalous 
dimensions in strong coupling. The resulting function can be organised 
in a perturbative series as
\begin{align}\label{eq:resum-exponent-expansion}
G_{\Nb}^I = \ln \Nb ~g_1^I(\omg) + g_2^I(\omg) + a_s ~g_3^I(\omg) + a_s^2~ g_4^I(\omg) + \cdots \,,
\end{align}
where $\omg = 2 \beta_0 a_s \ln \Nb$. Note that in the context of 
resummation $a_s \ln \Nb \sim \mathcal{O}(1)$. The successive terms in 
the above series along with the same for $g_{0,p}^I$ in \eq{eq:g0-expansion} 
define the resummation accuracies LL, NLL etc. The required exponents can 
be found in  \cite{Catani:2003zt,Moch:2005ba,H.:2020ecd,Das:2019btv}.
The coefficients $g_{0,p}^I$ which contains the process dependent 
information can be extracted from the newly computed SV coefficients by 
just transforming all the $\delta(1-z)$ and plus distributions into the 
Mellin space and dropping any $N-$dependent terms. Note that 
these coefficients differ significantly compared to the universal
scenario due to the fact that they now explicitly depend on the couplings.
However in the limit when these couplings ($k_q,k_g$) are same, they
should reproduce the exact coefficients for the universal case. 
This serves a crucial check for our computation. 
 We collect these newly 
computed coefficients up to the third order in the 
\app{app:g0-coefficients}.

The resummed expression in \eq{eq:resum-partonic} obviously does not 
contain any contributions which are not enhanced in the threshold limit 
particularly the regular contributions. These regular pieces from the quark 
or gluon initiated process or from other subprocesses are also important to 
describe the full phase space. A consistent way to include those 
contribution is generally performed through matching procedure as
\begin{align}\label{eq:matche-crosss-section}
\bigg[\frac{d\sigma}{dQ}\bigg]_{N^nLO+N^nLL} &= \bigg[\frac{d\sigma}{dQ}\bigg]_{N^nLO} +
\sum_{ab\in\{q,\bar{q},g\}}
 \frac{d\hat{\sigma}_{LO}}{dQ}  \int_{c-i\infty}^{c+i\infty} \frac{dN}{2\pi i} (\tau)^{-N} 
 \nn\\
& \delta_{ab}f_{a,N}(\mu_f^2) f_{b,N}(\mu_f^2) 
\times \bigg( \bigg[ \frac{d\hat{\sigma}_N}{dQ} \bigg]_{N^nLL} - \bigg[ \frac{d\hat{\sigma}_N}{dQ} \bigg]_{tr}     \bigg) \,.
\end{align}
The first term on the right hand side is the exact fixed order 
cross-section at N$^n$LO. The terms in the parentheses resums all the 
large logarithms avoiding any double counting from the fixed order counter 
part. Note that resummation is performed in Mellin space and finally one 
has to perform a Mellin-inversion. In particular one has to avoid the 
Landau pole while performing the inversion. This can be done using Minimal 
prescription \cite{Catani:1996yz} or Borel prescription 
\cite{Forte:2006mi,Abbate:2007qv}. We followed the former\footnote{For a 
detailed understanding of Borel prescription and comparison with Minimal 
ones, one can follow {\it eg.} \cite{Bonvini:2012sh}.} approach and fixed 
the contour accordingly.

Finally we conclude this section with the note that although generally 
the large Mellin-N logarithms are exponentiated, it is also possible to 
exponentiate some part or all $g_{0,p}^I$ \cite{Magnea:1990zb,Aybat:2006mz,
Dixon:2008gr,Bonvini:2014joa,Das:2019btv, H.:2020ecd}. However in this 
article we sticked to the standard approach and studied its effect at the 
LHC. 

\section{Numerical Results}\label{sec:numerics}
%
We now turn to the discussion on resummed numerical result up to NNLO+NNLL accuracy 
in QCD for dilepton production via spin-2 
particle at the LHC with non-universal coupling of spin-2 particle 
with gauge 
bosons and fermions. We remind the 
coupling strength of spin-2 particles to bosons 
through $\kappa_G = \sqrt{2} k_g/\Lambda$ and to fermions through
$\kappa_Q = \sqrt{2} k_q/\Lambda$. 
Here, $\Lambda$, the scale of the theory, is of the order 
a few TeV and we choose $0 < (k_q, ~k_g) < 1$. 
For this analysis we use {\tt PDF4LHC15} \cite{Butterworth:2015oua} parton distribution functions 
(PDF) throughout from {\tt{LHAPDF}} \cite{Buckley:2014ana} subroutine
unless otherwise stated. 
For the fixed order as well as resummed computation, 
we convolute NLO(NNLO) level partonic 
coefficient functions with 
NLO(NNLO) PDF.
The corresponding strong
coupling constant $\alpha_s(\mu_r^2)$ is also provided by 
{\tt{LHAPDF}} subroutine and for convenience we have defined
$a_s(\mu_r^2) = \alpha_s(\mu_r^2)/(4\pi)$.
The fine structure constant is taken to be $\alpha_{em} = 1/128$ and 
the weak mixing angle is $\sin^2\theta_w = 0.22343$. 
Here we present the results for $n_f = 5$ flavors in the massless 
limit of quarks. Our default choice for  the center of 
mass energy of the LHC is $E_{\rm CM}=13$ TeV. 
Except for the study of scale 
variations, we set renormalization ($\mu_r$) and 
factorization ($\mu_f$) scales equal to the dilepton invariant mass, 
{\it i.e.}
 $\mu_r = \mu_f = Q$. 
Our default choice of the cut-off scale of the theory 
is $\Lambda = 3$ TeV.

\begin{figure}[ht]
        \centerline{
        \includegraphics[width=7.0cm, height=7.0cm]{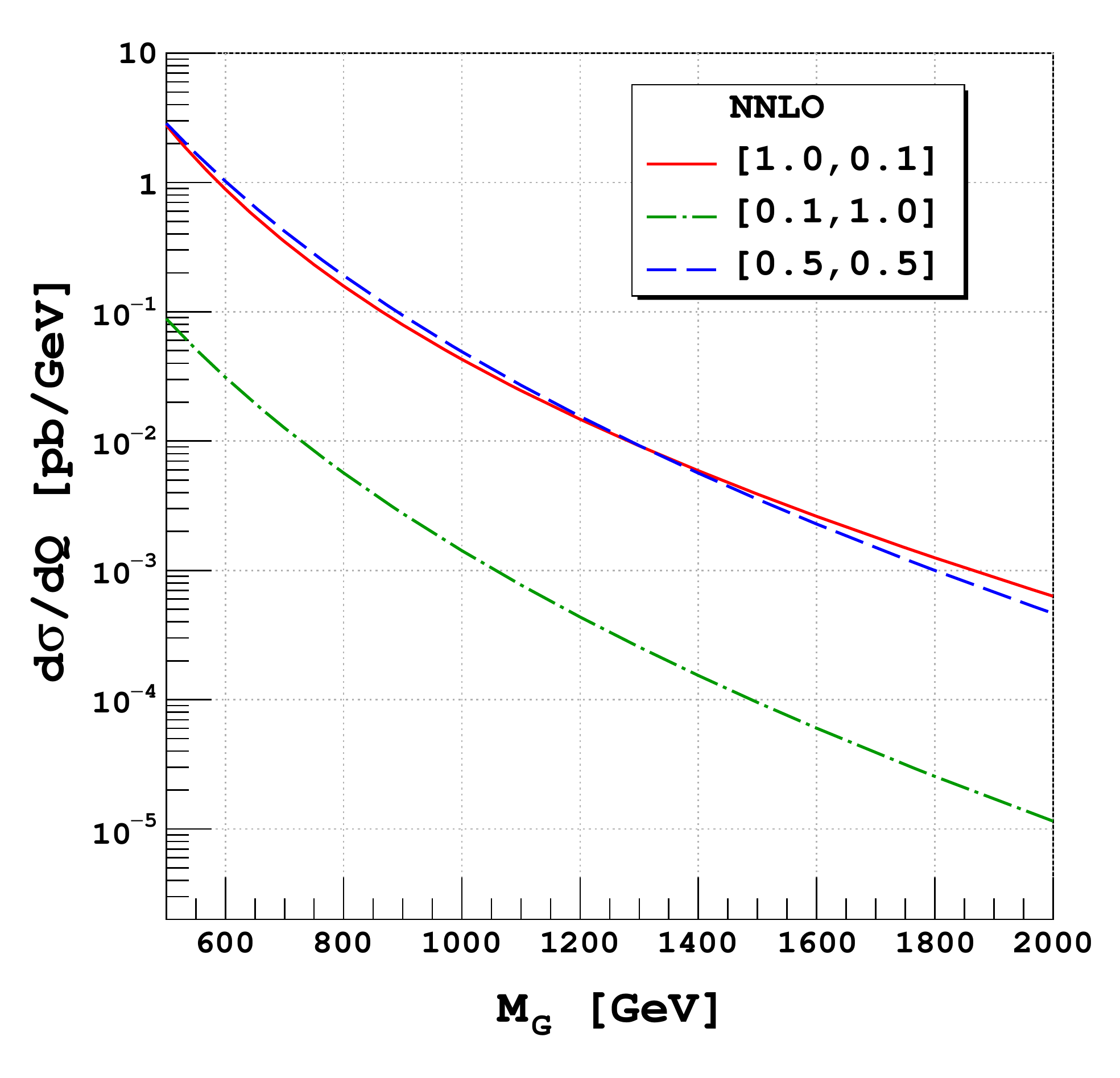}
        \includegraphics[width=7.0cm, height=7.0cm]{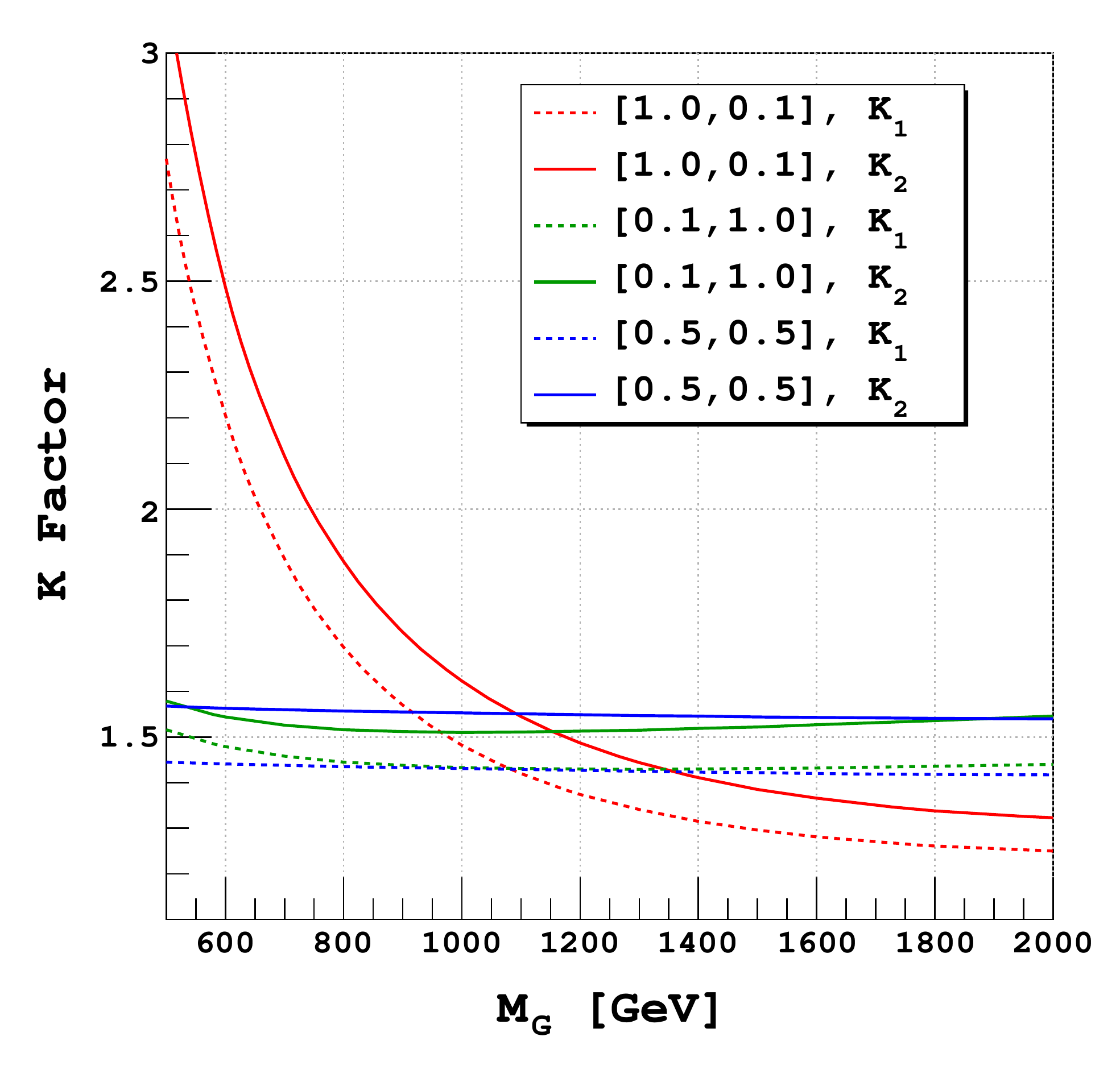}
        }
        \vspace{-2mm}
        \caption{\small{Invariant mass distribution of the dilepton for signal for different choice of $k_q$ and $k_q$ at $Q = M_G$ in NNLO QCD
               (left panel) and the corresponding NLO and NNLO K-factor as defined \eq{eq:fo_kfac} (right panel)}}
        \label{fig:fo_kfac}
\end{figure}

Before we present the resum result, we discuss some distinctive 
features of this model 
that are not observed for the case of universal couplings. 
For this, in \fig{fig:fo_kfac} left panel we 
present the dilepton invariant mass distribution at the resonance $Q=M_G$ 
for different choices of the model parameters $k_q$ and $k_g$.
For the results presented just in \fig{fig:fo_kfac}, we use NNLO PDF 
at all orders.
We observe that even at the resonance region, the cross section depends on the choice of  $k_q$ and $k_g$, 
which is not the case for the RS model where the height of the peak is 
independent of the corresponding coupling $c_0$ \cite{Das:2020gie}.
This particular nature can be understood from 
the parton level Born cross sections, \eq{eq:kq_kg_dependent_qqb}.
In \fig{fig:fo_kfac} right panel we present the corresponding K-factors 
($K_1,K_2$) defined with respect to LO
as,
\begin{align}\label{eq:fo_kfac}
        K_{n} = \frac{d\sigma^{N^nLO}/dQ}{d\sigma^{LO}/dQ}\,.
\end{align}
Here we observe that for $(k_q,k_g) = (1.0,0.1)$ at 
low $Q=M_G$ values, 
the NLO $K$-factors $K_1$ can be as high as $3$ while the corresponding 
$K$-factors for the universal case are 
about $1.5$ \cite{Das:2020gie}.  Because $k_q$ is much larger than $k_g$, at LO one can expect 
dominant contributions from Drell-Yan like 
process while the Higgs-like process is suppressed because of smaller 
coupling $k_g$. For the same reason, at NLO the gluon emissions from
the underlying born level parton processes is also expected to give 
smaller contribution. 
However, at NLO the presence of additional subprocesses like $qg$ with large 
parton fluxes, give a significant positive contribution comparable to that
 of $q\bar{q}$ LO as discussed in the introduction.
However for $(k_q,k_g) = (0.1,1.0)$ and $(0.5, 0.5)$, the $qg$ subprocess contribution is very small and comparable to LO $q\bar{q}$ subprocess 
and is negative. 
\begin{figure}[h]
        \centerline{
        \includegraphics[width=8cm, height=7.0cm]{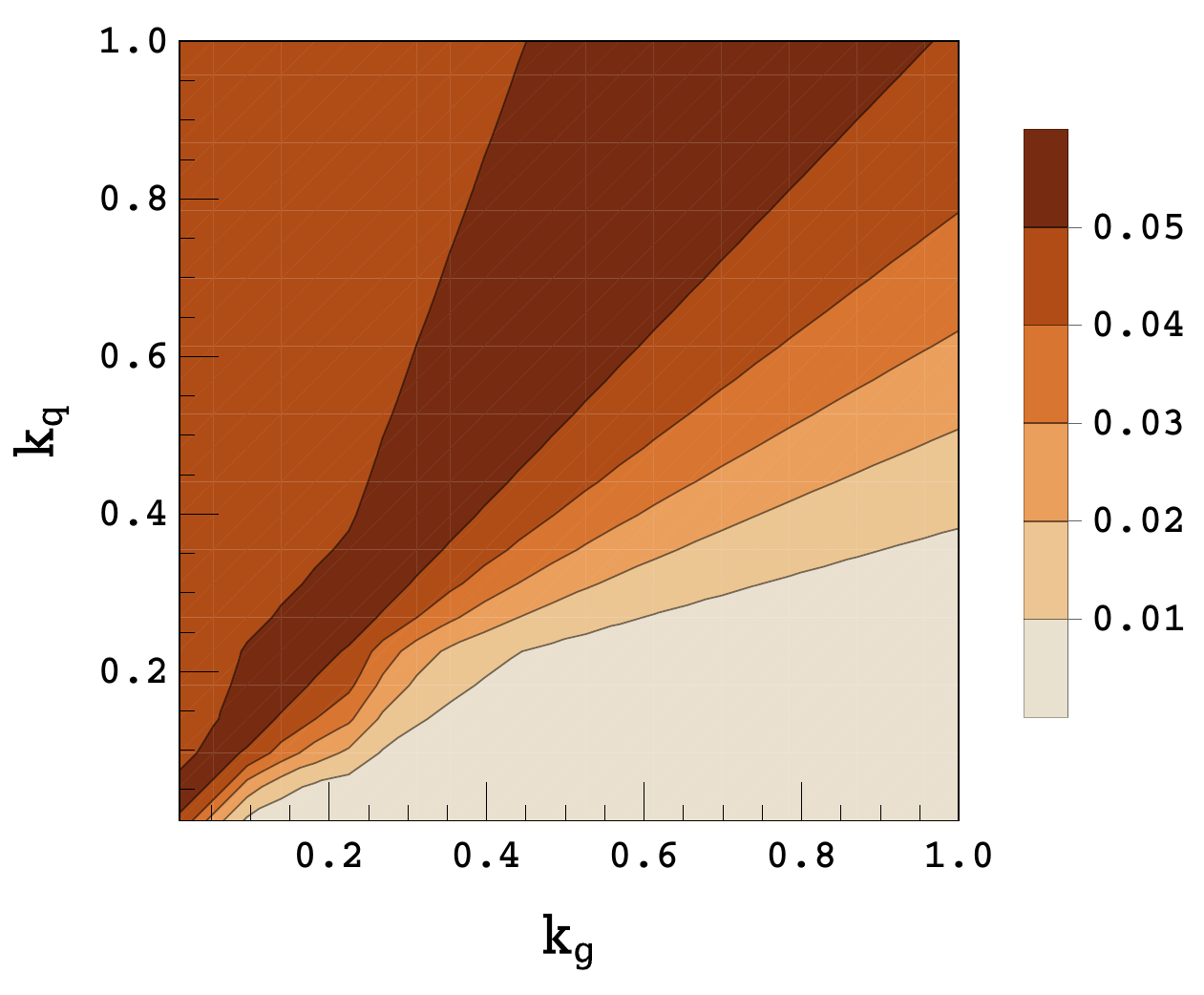}
        }
        \vspace{-2mm}
        \caption{\small{The contour plot of couplings $k_q$ and $k_g$
        for signal at NNLO with a fixed value of the spin-2 mass $M_G = 1$ TeV.
        The plot shows the cross-section (in $pb$) at the resonance $Q = M_G$.}}
        \label{fig:contourNNLO_kq_kg}
\end{figure} 
Because of this potentially large and model dependent contributions
 at NLO, it is necessary to include the QCD corrections at
 second order and beyond in the perturbation series.
We present the NNLO K-factors ($K_2$) in the right panel of \fig{fig:fo_kfac}
 and it can be seen that these second order contributions are smaller
than the corresponding NLO ones for different extreme choices of the non-universal couplings and confirms that at NLO all possible dominant contributions
are considered giving typical sizeable corrections from NNLO onwards. In this regard, it is convenient to study the higher order QCD corrections
in terms of the K-factors $R_{nm}$ defined with respect to NLO.
In the rest of the analysis, we present the results in terms of 
these K-factors $R_{nm}$.

Next, we present the dependence of model parameters $k_q$ and  $k_g$ on the FO 
cross sections at the
resonance for $Q=M_G = 1$ TeV through the contour plot
in  \fig{fig:contourNNLO_kq_kg}.
The cross section is large for larger $k_q$ values
 and becomes maximum 
around the universal line ($k_q \sim k_g$). 
The behaviour can be better understood from the dependence 
of born level partonic subprocesses 
on $k_q$ and $k_g$ as,
\begin{align}\label{eq:kq_kg_dependent_qqb}
	\Delta_{q\bar{q}}^{GR} \propto \frac{k_q^4}{C(Q^2) + [Ak_q^2 + Bk_g^2]^2} \quad , \qquad \quad \quad
	\Delta_{gg}^{GR} \propto \frac{k_q^2 k_g^2}{C(Q^2) + [Ak_q^2 + Bk_g^2]^2}\,.
\end{align}
Here the 
coefficients $A$ and 
$B$ contain the contributions  
for the decay of the spin-2 particle to fermions and bosons respectively 
and $C(Q^2) = (Q^2-M_G^2)^2$. At resonance where $C(Q^2) = 0$, the 
cross section will increase with increasing $k_q$ for a fixed $k_g$ 
and $k_g$ effect is mild here. 
For any given $k_q$, the cross sections are maximum when $k_g \lesssim k_q$.

\begin{figure}[ht]
        \centerline{
        \includegraphics[width=7.0cm, height=7.0cm]{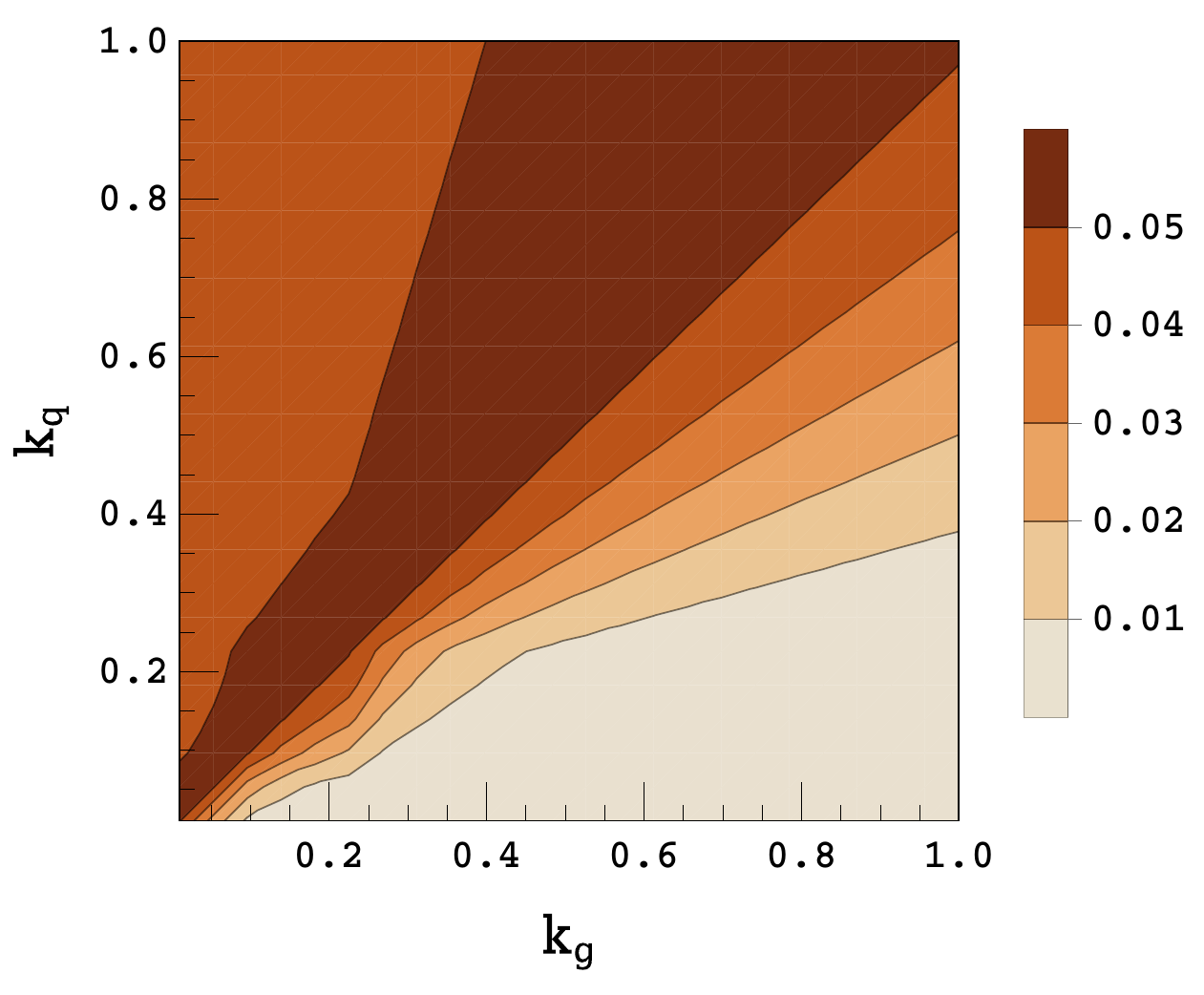}
        \includegraphics[width=7.0cm, height=7.0cm]{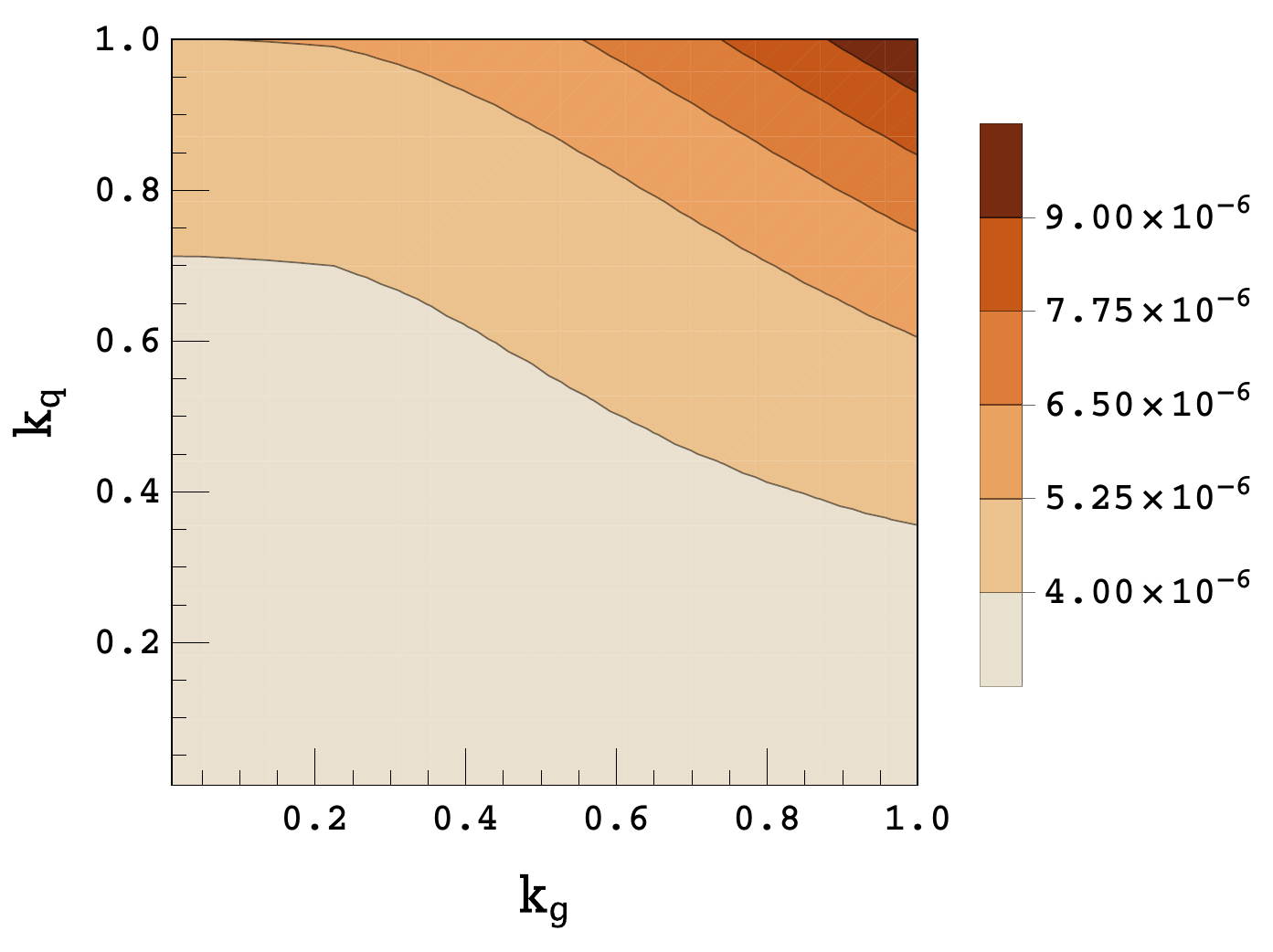}
        }
        \vspace{-2mm}
        \caption{\small{The contour plot of couplings $k_q$ and $k_g$
        for signal at NNLO+NNLL with a fixed value of the spin-2 mass $M_1 = 1$ TeV.
        The left plot shows the cross-section (in $pb/GeV$) at the resonance $Q=M_1$, the
        right plot shows the cross-section (in $pb/GeV$) away from the resonance at $Q=1.5$ TeV.}}
        \label{fig:contourNNLL_kq_kg}
\end{figure}

\begin{figure}[ht]
        \centerline{
        \includegraphics[width=7.0cm, height=7.0cm]{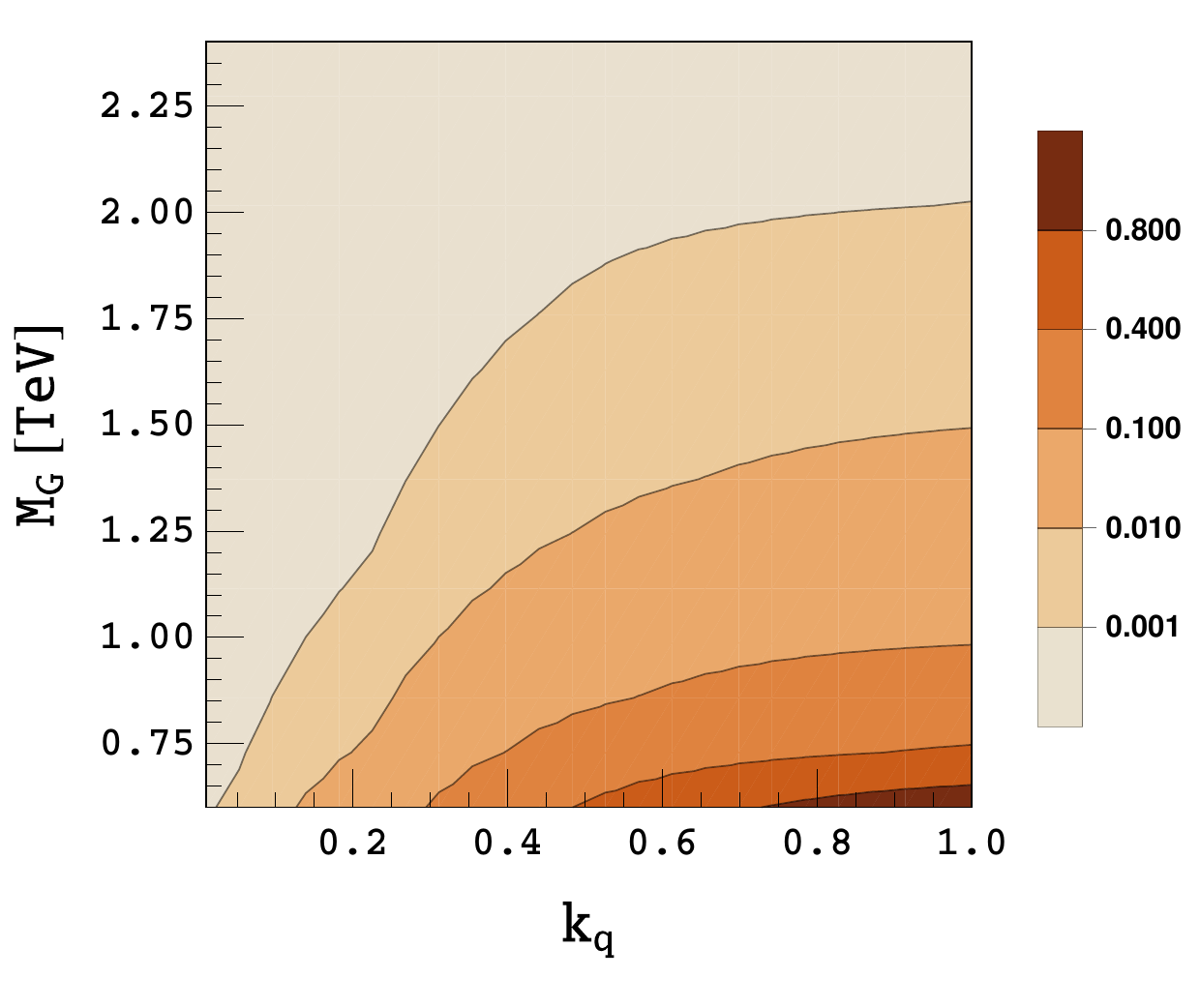}
        \includegraphics[width=7.0cm, height=7.0cm]{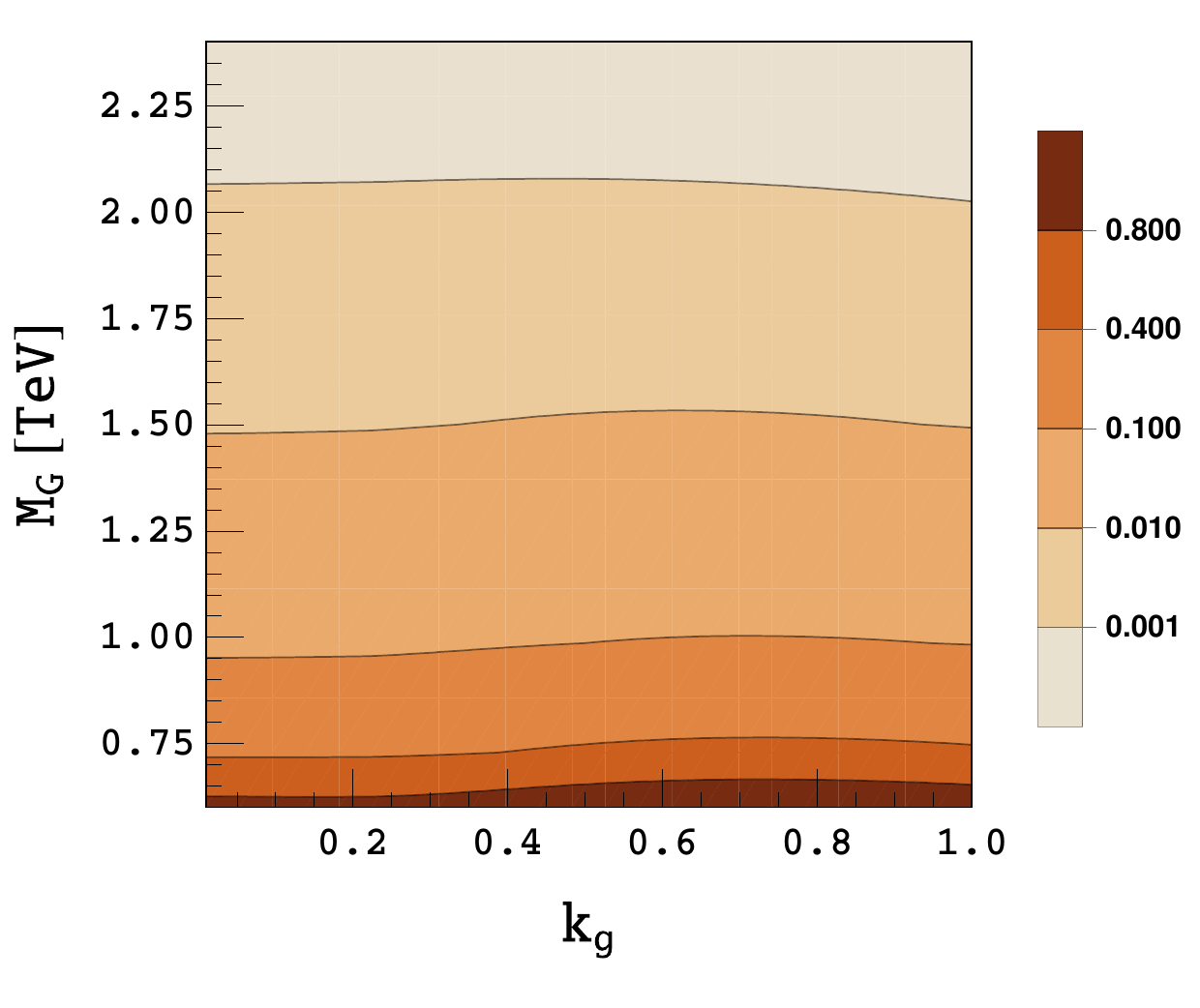}
        }
        \vspace{-2mm}
        \caption{\small{The contour plot for mass of the spin-2 
        particle with coupling $k_q$ and $k_g$ for signal at NNLO+NNLL 
        at $Q = M_G$.
        The left panel shows the cross section (in $pb$) for different 
        $M_G$ and $k_q$ region while $k_g = 1.0$ and the right panel  
        shows the cross section for $M_G$ and $k_g$ region with $k_q$ 
        fixed at $1.0$.}}
        \label{fig:contourNNLL_kqkg_M1}
\end{figure}

We now discuss the resummed effects for the non-universal couplings.
We have performed resummation up to NNLL accuracy and matched them with 
the fixed order NNLO level as described in \sect{sec:theory}.
In \fig{fig:contourNNLL_kq_kg}, we have shown the dependence of the cross 
section on $k_q$ and $k_g$ including the resummation effect at NNLL. 
In the left panel, we present the contour region at the resonance 
$Q=M_G=1$ TeV and in the right panel we
present the same but at the off-resonance region $Q=1.5$ TeV. 
Note that unlike the RS case,
 here we find non-negligible GR contribution away 
from the resonance as well.
The dependence on the model parameters is found
to be the similar as that observed in the FO case 
\fig{fig:contourNNLO_kq_kg}. However, the maximum cross section region for the
resummed results is wider.
We have also studied the cross-section regions for varying 
the spin-2 mass $M_G$  and either of the couplings while keeping 
the other coupling constant. 
In \fig{fig:contourNNLL_kqkg_M1}, we present such 
behaviour at the resonance region $Q=M_G$ to NNLO+NNLL accuracy 
for different values of $M_G$ and $k_q$ for a given 
$k_g=1.0$ in the left panel.  
We observe that with increasing $M_G$ the 
cross section is falling for a fixed $k_q$. This could be described due to the 
decreasing parton fluxes.
The cross section is increasing with $k_q$ for a fixed value of $M_G$ as 
discussed in \eq{eq:kq_kg_dependent_qqb}.
In the right panel, we present similar results but by varying $M_G$ and $k_g$ 
for a given value of $k_q=1.0$
and we observe the mild effect of $k_g$ on cross section when both $M_G$ and 
$k_q$ are fixed.

\begin{figure}[ht]
        \centerline{
        \includegraphics[width=8.0cm, height=7.0cm]{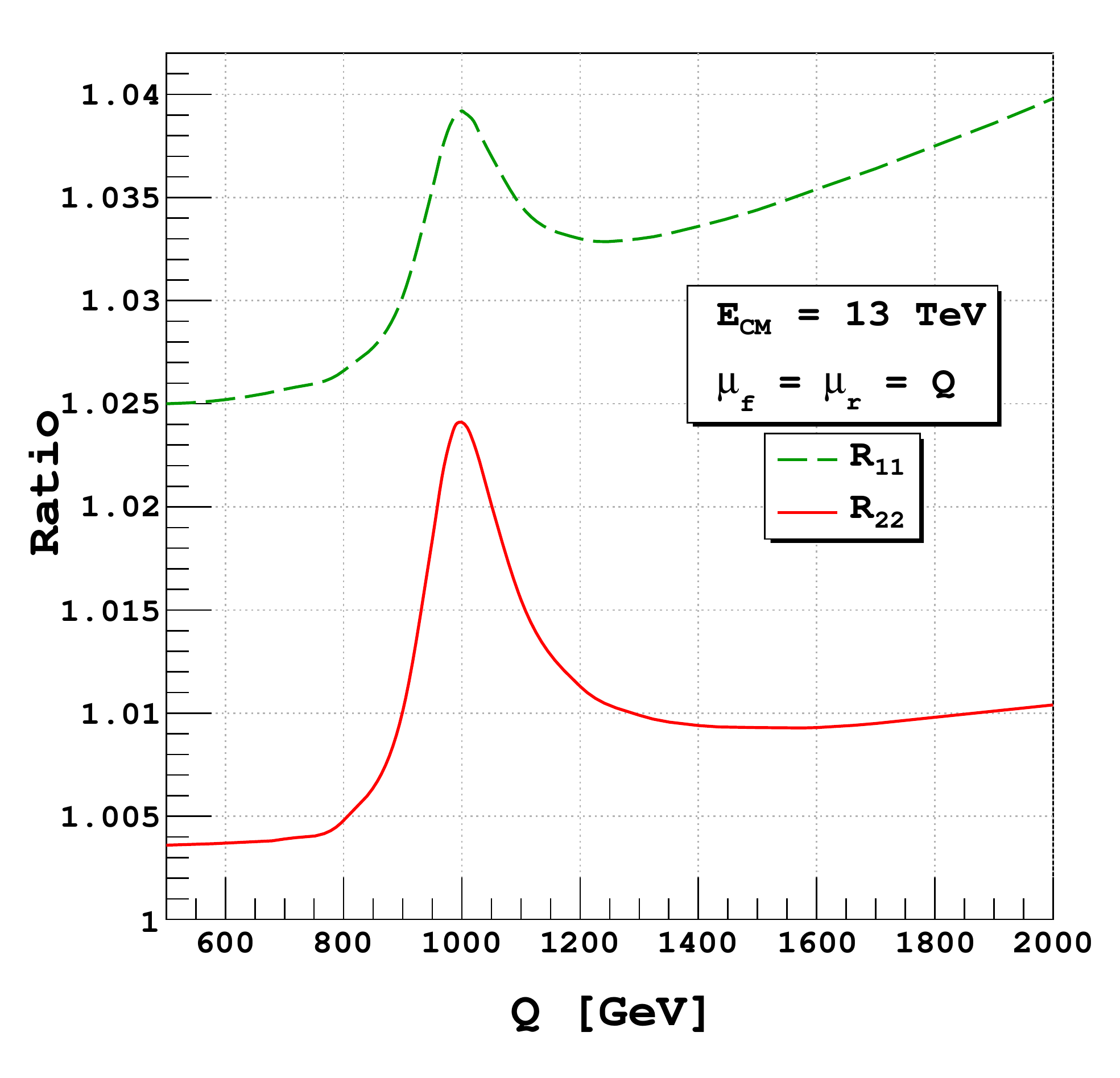}
        }
        \vspace{-2mm}
        \caption{\small{Comparison between fixed order results and 
        resummed results. The ratios are defined with respect to NLO
        as described in the text.}}
        \label{ratio1}
 \end{figure}
To quantify the resummation effect we define K-factor with respect to 
NLO as discussed before. We define,
\begin{align}\label{eq:resum_ratio}
        R_{nm} = \frac{d\sigma^{(N^nLO+N^nLL)}/dQ}{d\sigma^{(N^mLO)}/dQ}\,.
\end{align}
We present in \fig{ratio1}, the ratio of resum  results matched to 
the FO ones as given in \eq{eq:resum_ratio}. 
We notice that there is around $4\%$ enhancement due to NLL resummation 
over the NLO FO result at the resonance region
while the enhancement due to NNLL resummation is around $2.5\%$ over the 
NNLO results.
In general we notice that in the region $Q>M_G$, the resummation effects 
keep increasing with $Q$ and are dominant
at NLO+NLL level, whereas they are almost constant and are about 
$1\%$ at 
NNLO+NNLL compared to NNLO. 

\begin{figure}[ht]
        \centerline{
        \includegraphics[width=7.0cm, height=7.0cm]{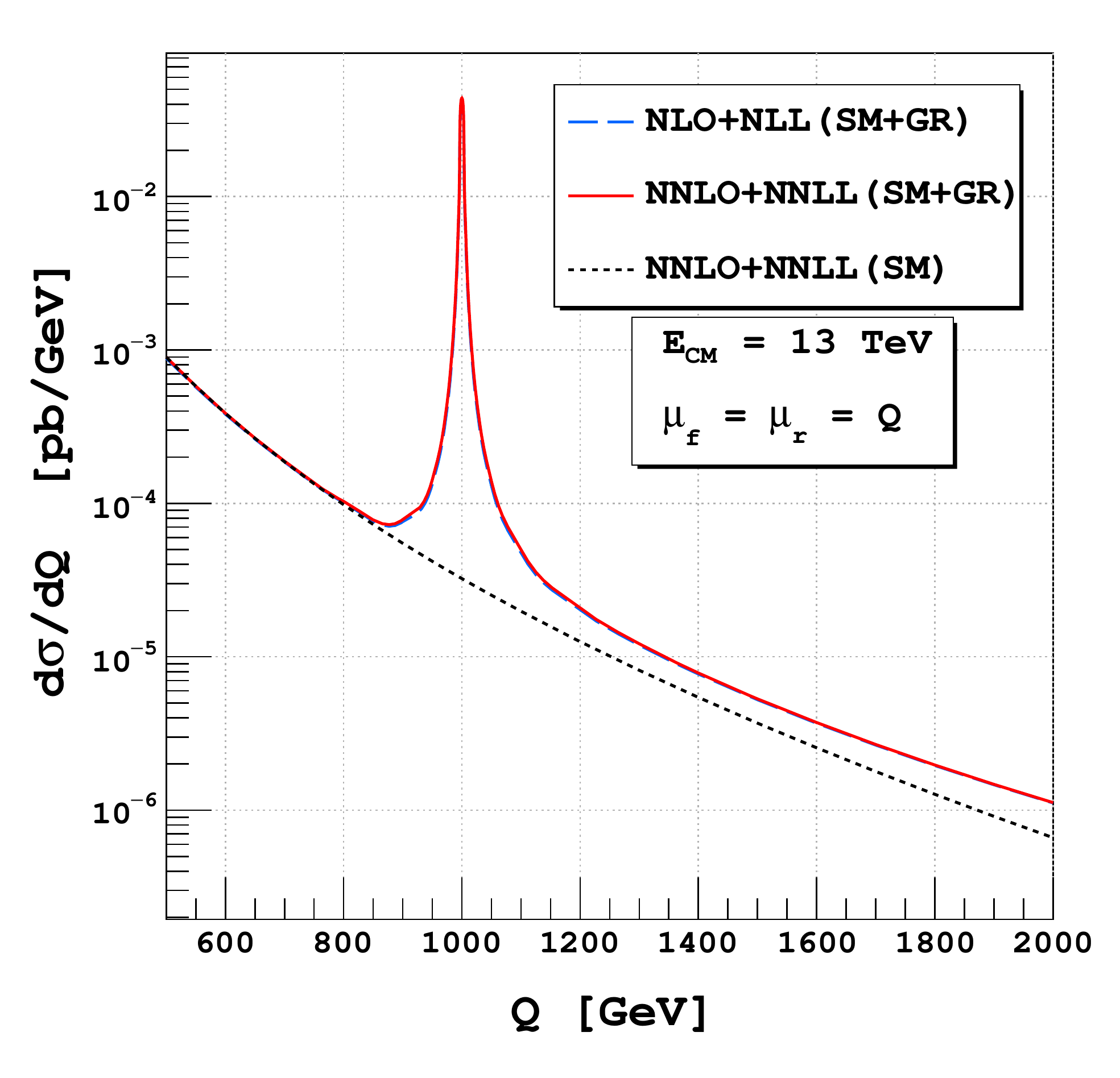}
        \includegraphics[width=7.0cm, height=7.0cm]{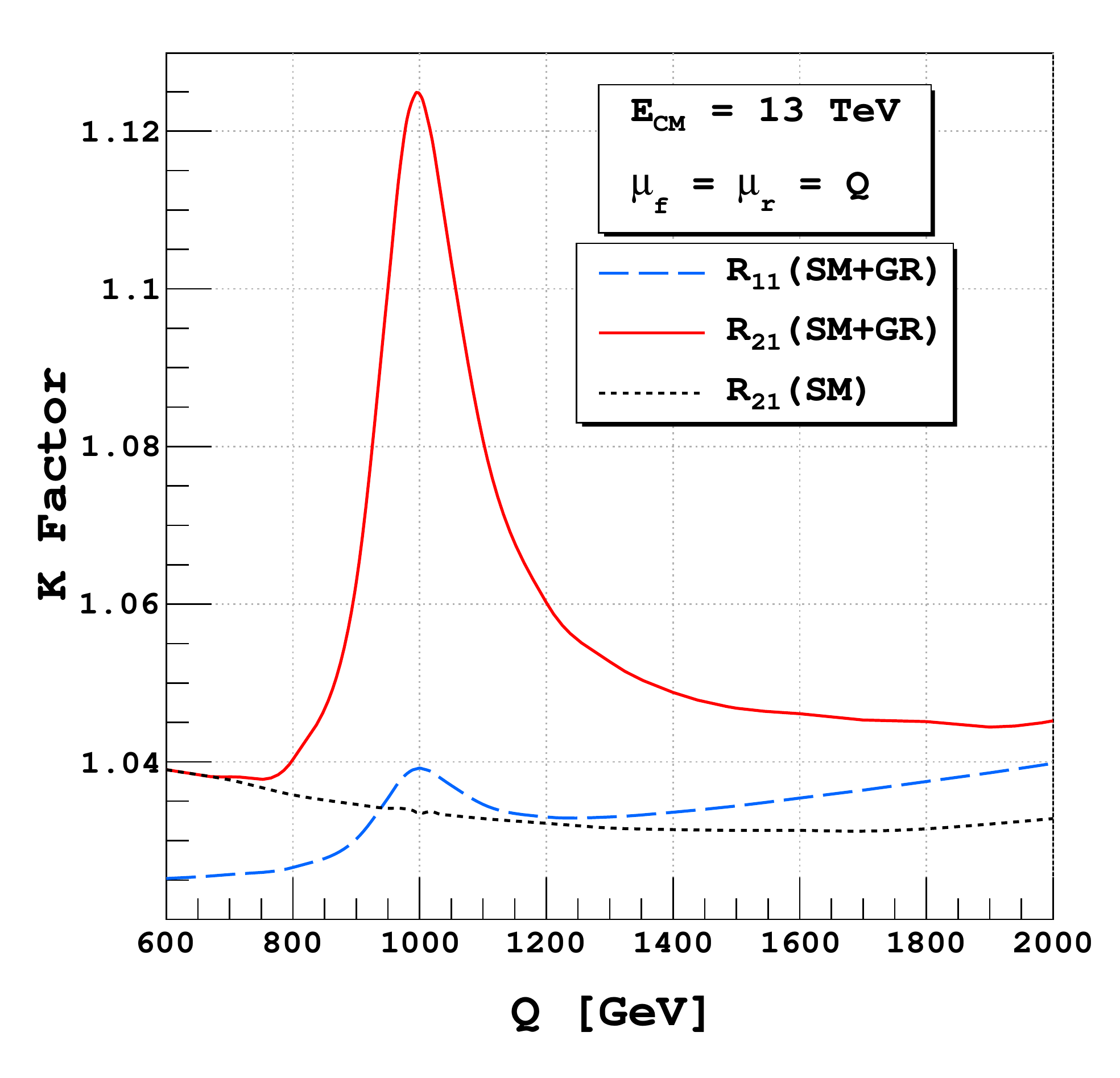}
        }
        \vspace{-2mm}
        \caption{\small{Dilepton invariant mass distributions 
        are presented 
         to NNLO+NNLL QCD for signal (left panel) and the corresponding 
         $K$-factors (right panel).}}
        \label{sig_inv_res}
\end{figure}
\begin{figure}[ht]
        \centerline{
        \includegraphics[width=7.0cm, height=7.0cm]{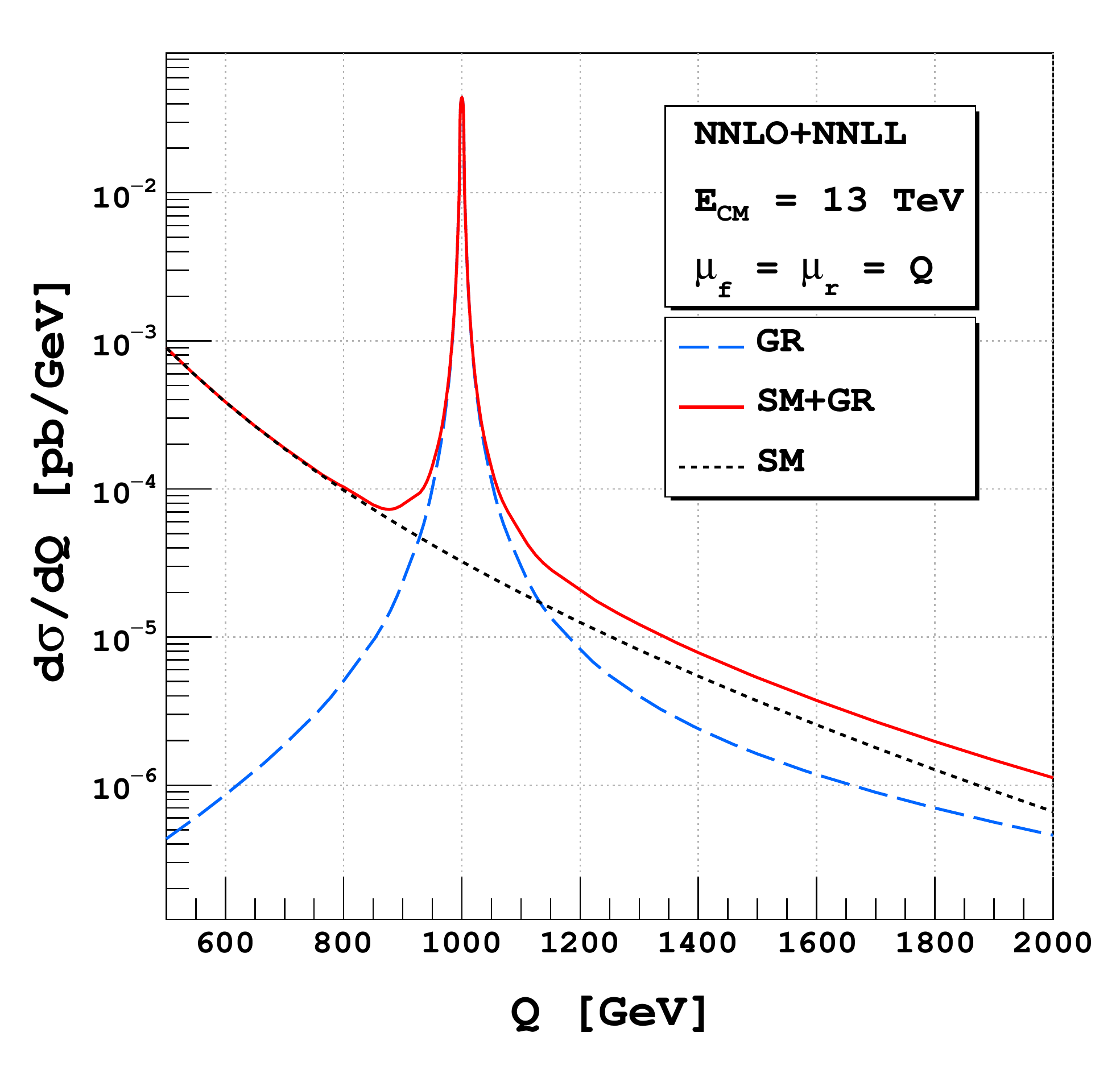}
        \includegraphics[width=7.0cm, height=7.0cm]{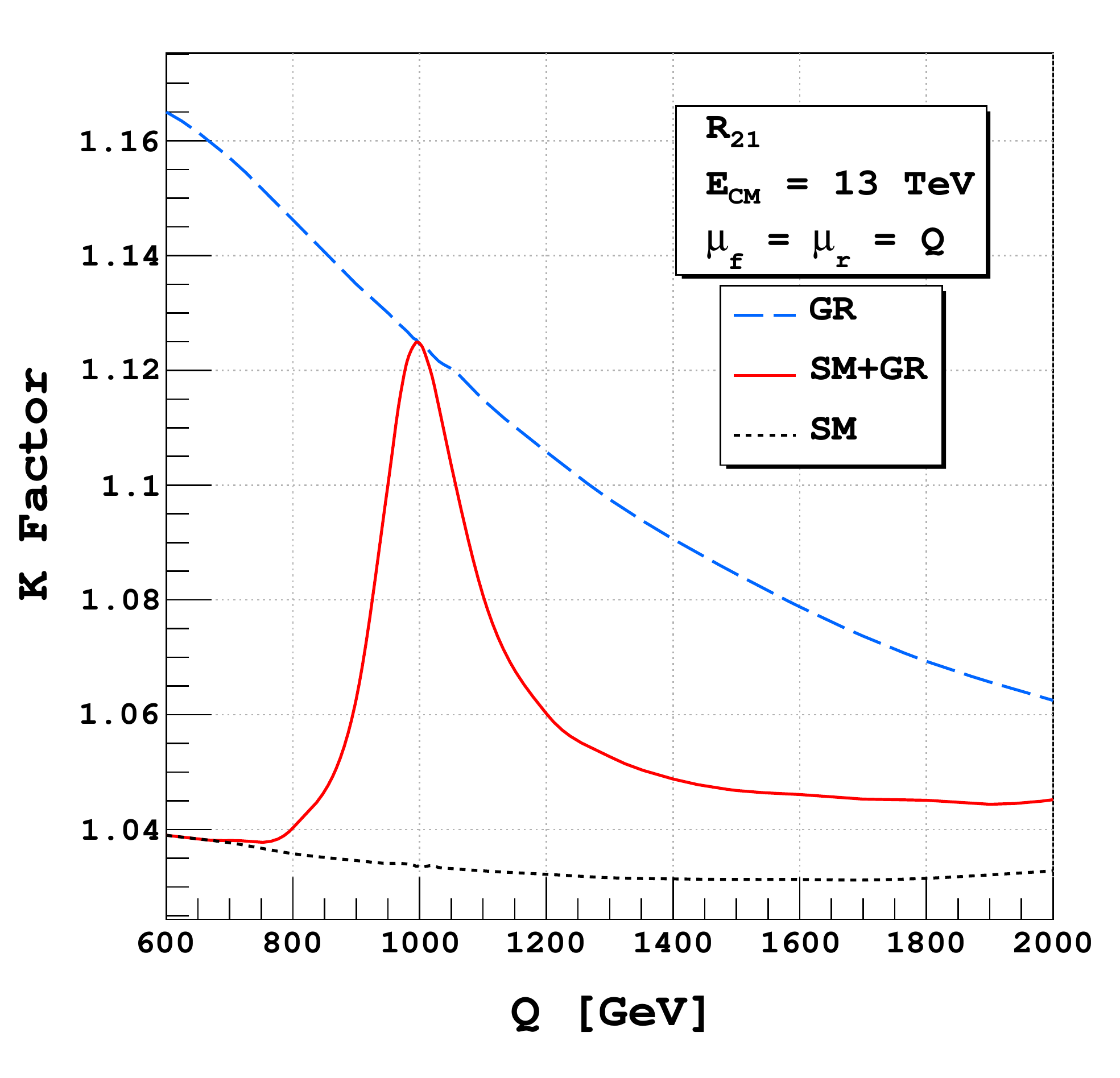}
        }
        \vspace{-2mm}
        \caption{\small{Invariant mass distribution of dilepton for the SM, GR model and for the signal (left)
        and their corresponding K-factors with respect to NLO (right)}}
        \label{model_res}
\end{figure}
\begin{figure}[ht]
        \centerline{
        \includegraphics[width=8.0cm, height=7.0cm]{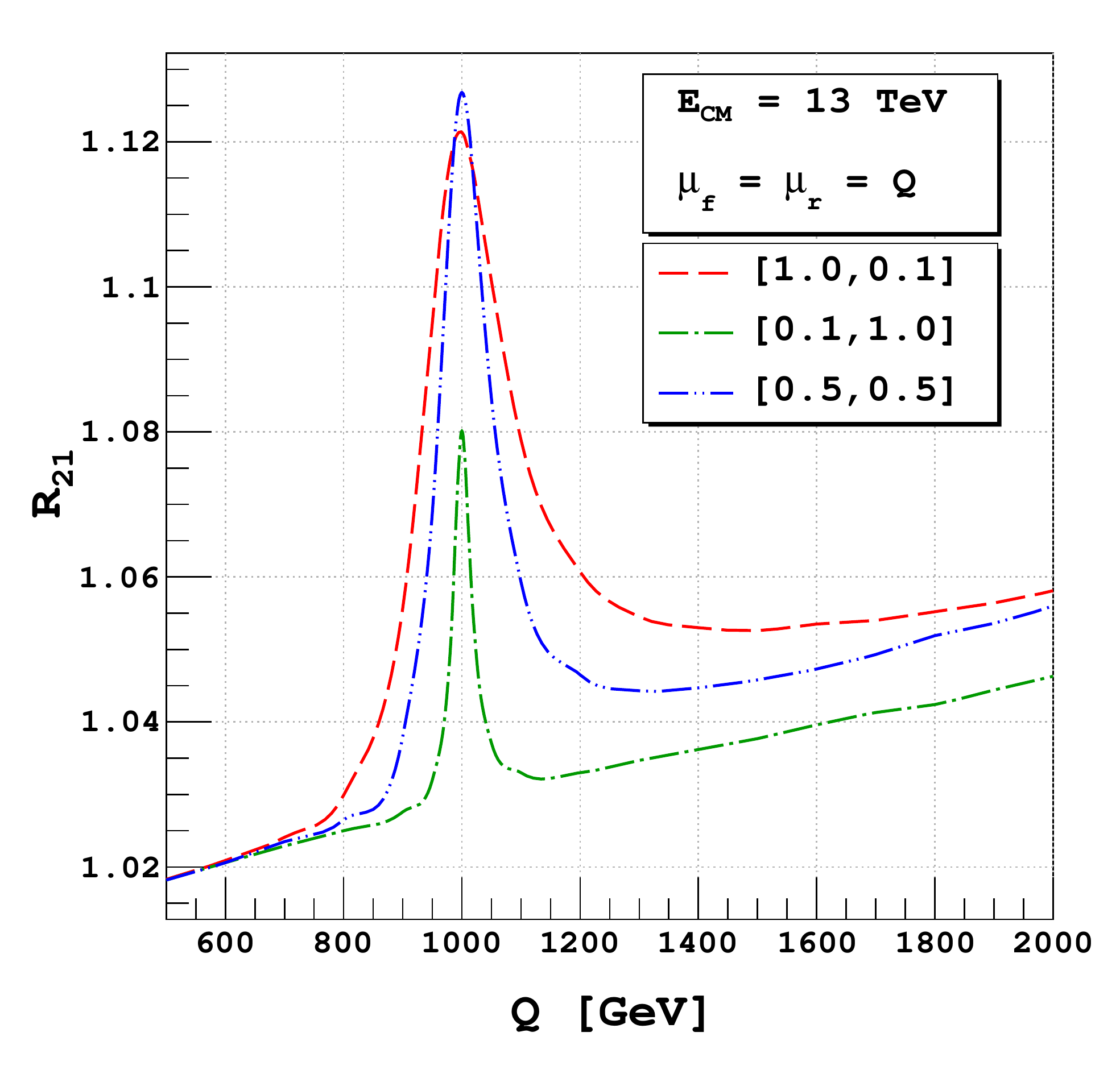}
        }
        \vspace{-2mm}
        \caption{\small{K-factor of signal with respect to NLO at NNLO+NNLL level 
        for different choice of $k_q$ and $k_g$}}
        \label{comparison}
\end{figure}

In \fig{sig_inv_res}, we present the invariant mass distribution 
to NNLO+NNLL accuracy in the left panel 
and the corresponding $K$-factors in the right panel for the default 
choice of model parameters. 
We observe a significant enhancement in the cross section from lower 
order to higher order.
In order to study different contributions coming from SM, pure GR and 
the signal, we
present  \fig{model_res}. 
We observe that the contribution of gravity in the invariant mass 
distribution is negligible for $Q < 900$ GeV.
However, after the resonance the gravity contribution is also
prominent which is in contrast to RS scenario as mentioned earlier.
In the right panel we present the corresponding 
$K$-factors for the SM, pure gravity and the signal. 
At low $Q$ value the signal $K$-factor is equal to that of SM. 
At the resonance, most of the 
signal contributions are coming from the gravity, therefore, 
the signal K-factor is almost equal to that of gravity.
In \fig{comparison}, we show the NNLO+NNLL $K$-factors for different
choices of couplings. It can be seen that the higher cross-section is 
achieved when $k_g \lesssim k_q$,

\begin{figure}[ht]
        \centerline{
        \includegraphics[width=7.4cm, height=5.0cm]{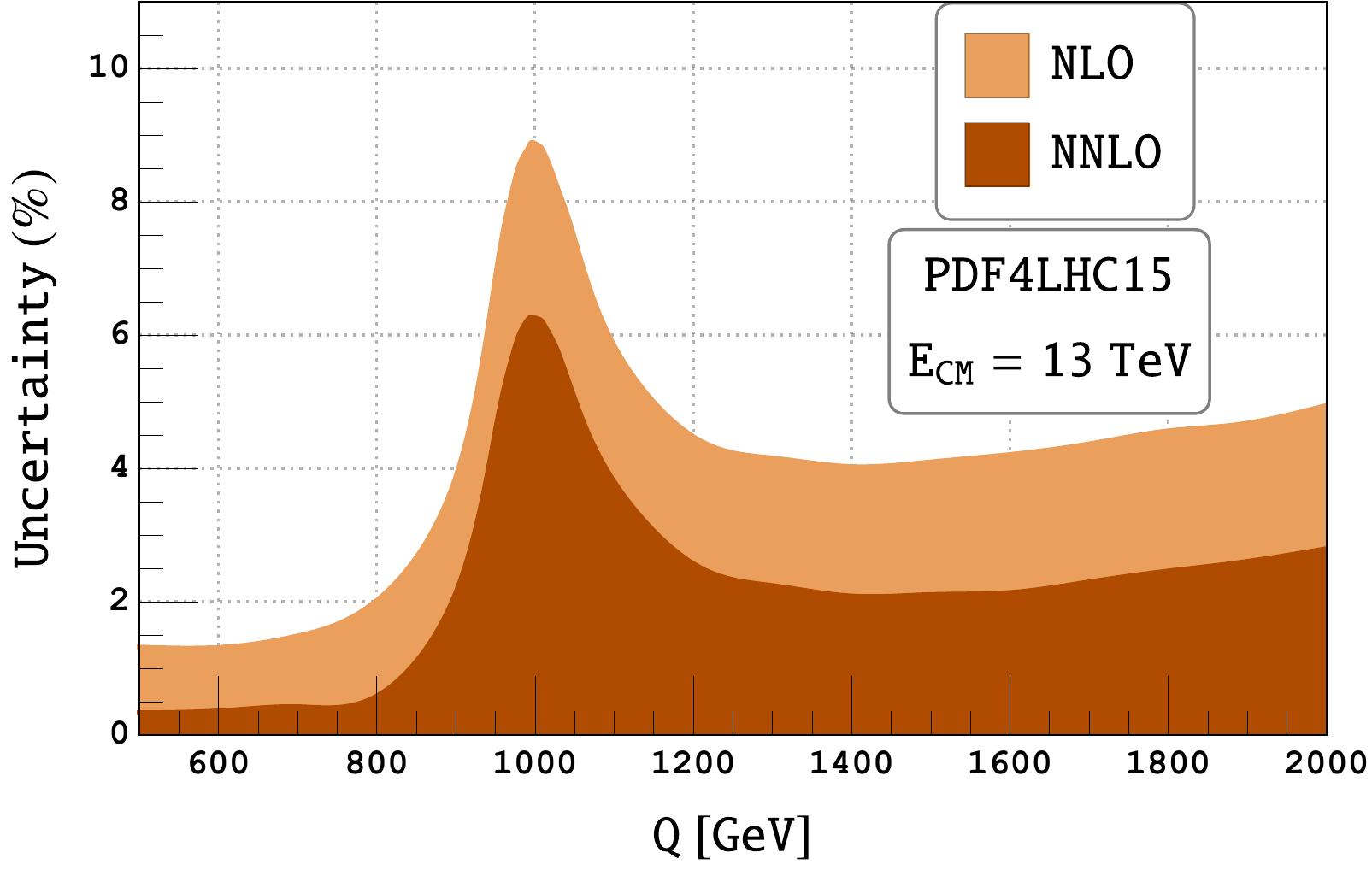}
        \includegraphics[width=7.4cm, height=5.0cm]{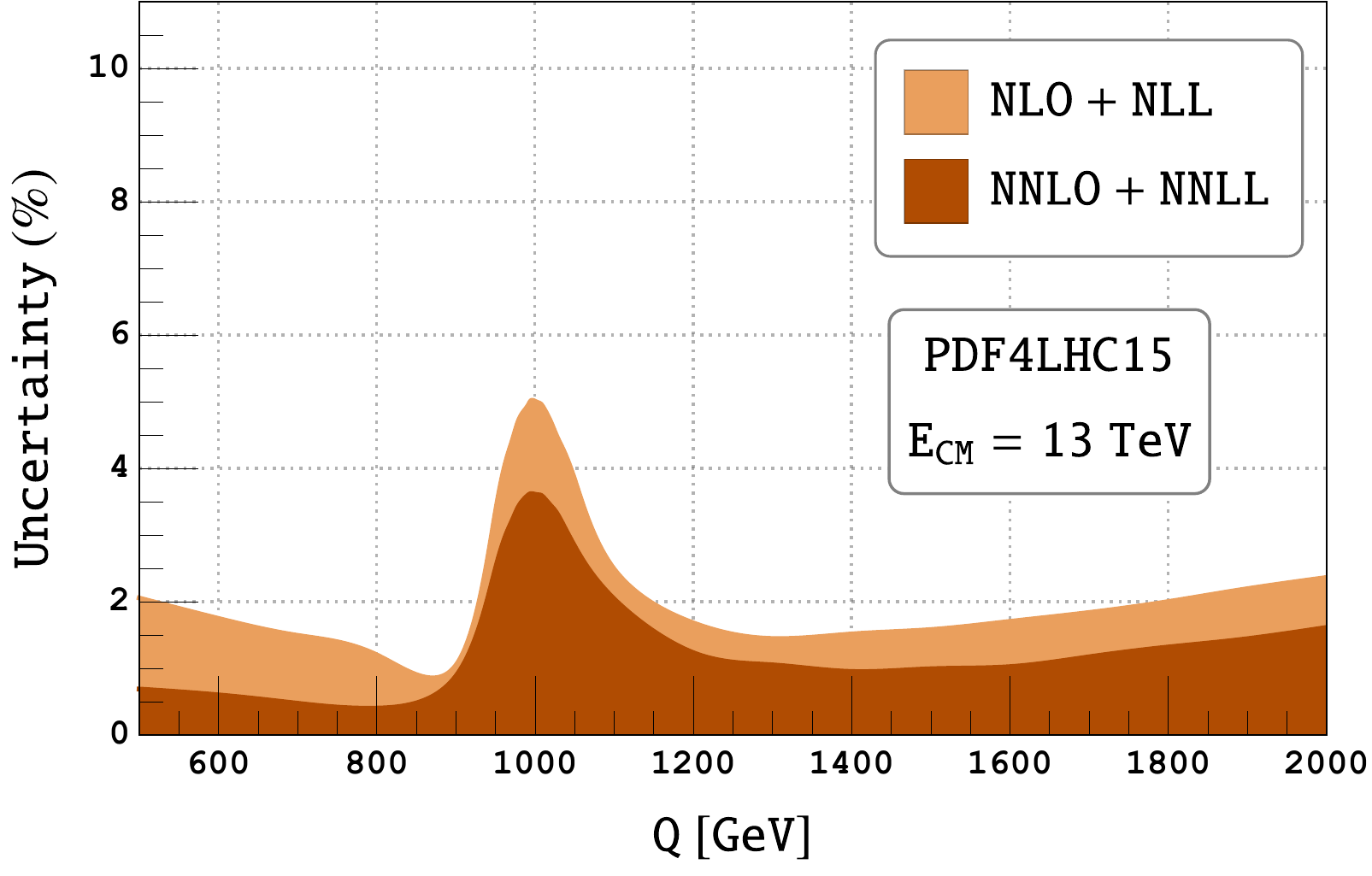}
        }
        \vspace{-2mm}
                \caption{\small{The 7-point scale variation in the signal 
                in the range $(\mu_r,\mu_f) \in (1/2,2)$ is 
                shown up to NNLO and NNLO+NNLL for the dilepton 
                invariant 
                mass distribution}}
        \label{7points_uncertainty_res}
\end{figure}

We also estimate various theoretical uncertainties in our analysis. 
We first consider the uncertainties due 
to the unphysical scales $\mu_r$ and $\mu_f$. 
To quantify these uncertainties, we use the conventional
canonical 7-points scale 
variations by varying $\mu_r$ and $\mu_f$ 
simultaneously from $Q/2$ to $2Q$ subject to the constraint that the 
ratio of unphysical scales should not be than $2$ and taking the maximum
absolute deviations. We put the following constraints on the variation
in order to remove extreme combinations,
\begin{align}\label{eqscale}
        \Big|\text{ln}\frac{\mu_r}{Q}\Big| \leq \text{ln }2, \quad \Big|\text{ln}\frac{\mu_f}{Q}\Big| \leq \text{ln }2,
        \quad \Big|\text{ln}\frac{\mu_r}{\mu_f}\Big| \leq \text{ln }2\,.
\end{align}     
In \fig{7points_uncertainty_res}, we present these 
 scale uncertainties both in the fixed order results (left panel) 
as well as in the resummed results (right panel). 
At NNLO this uncertainty is estimated to be around $\pm 0.6\%$ for $Q<M_G$, 
around $\pm 6.3\%$ at the resonance region $Q=M_G$ and is about $\pm 2.0\%$ for $Q>M_G$.
However, the corresponding scale uncertainties for the signal at NNLO+NNLL accuracy
found to get significantly reduced, respectively, to around $\pm 0.4\%$, $\pm 3.6\%$ and $\pm 1.0\%$. 
We have also estimated the uncertainties only due to the remormalization scale $\mu_r$ by varying it
from $Q/2$ to $2Q$ and keeping $\mu_f = Q$ fixed. We observe a significant reduction in the renormalization scale uncertainty 
from $\pm 1.4\%$ at NNLO to about $\pm 0.5\%$ at NNLO+NNLL accuracy. 
The corresponding factorization scale uncertainties
obtained by varying $\mu_f$ from $Q/2$ to $2Q$ and 
keeping $\mu_r=Q$ fixed are found to get reduced from $5\%$ at NNLO to
$2.6\%$ at NNLO+NNLL accuracy at the resonance.

Apart from the unphysical scale uncertainties, we also estimate 
the uncertainties in the fixed order NNLO result as well as the 
resummed NNLO+NNLL predictions coming from non-perturbative PDFs. 
We have estimated the intrinsic PDF uncertainty for the default
 choice of {\tt PDF4LHC15} PDFs using the recommendation in
\cite{Botje:2011sn},
 and present these results in tab.(\ref{table1}).
 We observe that the PDF uncertainty is increasing with invariant mass ($Q$).
 However, there is no significant improvement in the 
 PDF uncertainty after inclusion of 
 threshold logarithms. 
Furthermore, we also compute the cross sections at NNLO+NNLL accuracy for the central set $i=0$ of 
different PDF groups namely, 
{\tt MMHT2014nnlo68cl} \cite{Harland-Lang:2014zoa}, 
{\tt CT14nnlo} \cite{Dulat:2015mca}, {\tt ABMP16\_5\_nnlo} \cite{Alekhin:2013nda} and 
{\tt NNPDF31\_nnlo\_as\_0118} \cite{Ball:2017nwa} at resonance $Q=M_G=1$ TeV and 
at $Q = 1500$ GeV. 
The corresponding results obtained are found to be 
different (as much as $3\%$)  
in some cases than the our default choice PDF set.

%
\begin{table}[h!]
	\begin{center}
{\scriptsize
		\resizebox{15.0cm}{0.65cm}{
	 		\begin{tabular}{|c|c|c|c|c|c|c|}
\hline
 $Q (GeV)$ & $500$ & $1000$ & $1500$ & $2000$  & $2500$ \\
\hline
NNLO & $8.95\times 10^{-4}\pm 2.0\%$ & $4.28\times 10^{-2}\pm 2.2\%$ & $5.27\times 10^{-6}\pm 3.4\%$ & $1.11\times 10^{-6}\pm 4.2\%$ & 
				$3.37\times 10^{-7}\pm 5.1\%$ \\
\hline
NNLO+NNLL & $8.98\times 10^{-4}\pm 1.9\%$ & $4.39\times 10^{-2}\pm 2.2\%$ & $5.32\times 10^{-6}\pm 3.3\%$ & $1.12\times 10^{-6}\pm 4.2\%$ &
				$3.41\times 10^{-7}\pm 5.0\%$ \\
\hline
\end{tabular}
 }
		\caption{\small{Intrinsic PDF uncertainties
		 for {\tt PDF4LHC15} at $13$ TeV LHC}} 
\label{table1}
 }
 \end{center}
\end{table}

%
Although the main focus of our phenomenological study is on
the threshold resummation, we  have also studied
the soft plus virtual (SV) corrections at third order in QCD, N$^3$LO$_{\rm SV}$.
We have computed these third order SV coefficients using the three-loop forms
factors \cite{Ahmed:2016qjf} and exploiting the universality of soft radiations.
The new results are presented in \app{app:sv-coefficints}. 
We finally then give a numerical
estimate of these third order corrections by using the running strong coupling 
constant at $4$-loop \cite{vanRitbergen:1997va} level.
We use the same {\tt PDF4LHC15} NNLO set at this order and 
find that
the third order SV result contributes an additional $1\%$ to the 
NNLO result at the resonance region.
The renormalization scale uncertainty reduces to $\pm 0.2\%$ at resonance for
a canonical variation\footnote{Note that it is also possible to
estimate theoretical scale uncertainty through a probabilistic 
description as in \cite{Cacciari:2011ze,Bonvini:2020xeo}. 
However we refrain from this study in this article.}
within $\mu_r \in \{1/2,2\}Q$ keeping $\mu_f = Q$ fixed. 
However for a proper estimation of third order QCD corrections
(particularly in the region away from threshold region {\it viz.} in
the lower invariant mass region) one needs to include the regular terms
pieces as well as PDF at the third order. 
\begin{figure}[ht]
    \centerline{
    \includegraphics[width=10.5cm, height=7.0cm]{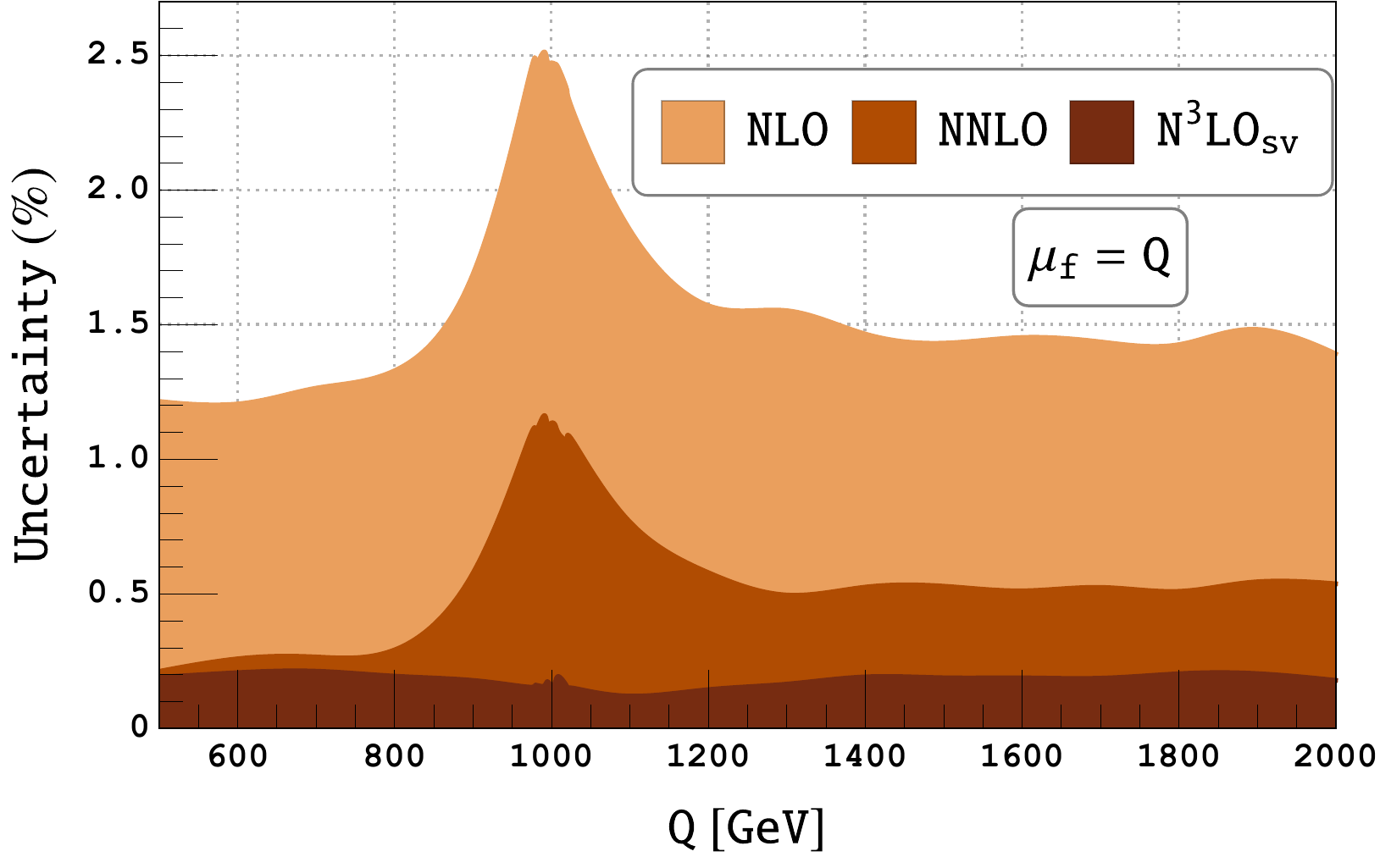}
    }
    \vspace{-2mm}
            \caption{\small{The uncertainty due to $\mu_r$ scale 
            variation around the central scale $Q$ at
             N$^3$LO$_{\rm SV}$ for 
            fixed $\mu_f$ = $Q$
            for signal}}
    \label{eq:mur_variation_n3lo}
    \end{figure}

%
\section{Summary}\label{sec:conclusion}
We have studied a detailed phenomenology for a generic spin-2 particle 
production at the 13 TeV LHC. We assumed non-universal coupling of
the spin-2 particle with the SM fields. This needs additional UV 
renormalization for individual operators. After performing UV 
renormalization of the form factors, we have computed the new 
 SV coefficients at the third order.
From these coefficients, we were able to extract the process dependent
coefficients needed for threshold resummation. 
Using the universal threshold exponent, which are already available
in the literature as well as the newly computed process dependent
coefficients in this article, we find all the ingredients to perform resummation
up to NNLL accuracy. We observe a better
perturbative convergence  after inclusion
of these threshold logarithms.  Compared to the fixed order, the
cross-section increases by $2.5\%$ at the NNLL level at the resonance. 
We show that inclusion of these threshold 
logarithms are indeed important in taming the theoretical uncertainties 
to as small as $3.6\%$ near the resonance.
We also discuss the impact of the third order SV coefficients and
the corresponding renormalization scale uncertainty.
We stress that the $K$ factors in the non-universal case strongly
 depend on the higher order corrections and a naive scaling
 from LO will heavily undermine the correct result at NLO(NNLO) level.
  Our mass-dependent
 $K$ factors are thus expected to be useful in the search of such spin-2 
 resonances at the LHC. 
 \section*{Acknowledgements}
 We would like to thank P. Banerjee and P. K. Dhani for useful discussion 
 and providing three loop GR form factor \cite{Ahmed:2016qjf} in electronic 
 format. The research of G.D. was supported by the Deutsche 
 Forschungsgemeinschaft (DFG, German Research Foundation) under grant  
 396021762 - TRR 257 (``{\it Particle Physics Phenomenology after the 
 Higgs Discovery}''). 
 \appendix
\section{Anomalous Dimensions}\label{app:anomalous-dimensions}
The UV anomalous dimensions appearing in Eq. (\ref{eq:uv-renorm-const}) 
can be written in terms of renormalized strong coupling as 
\begin{align}\label{eq:uv-renorm-expansion}
\gamma_{IJ} = \sum_{i=1}^{\infty} a_s(\mu_r)^i \gamma_{IJ}^{(i)} \,.
\end{align}
Up to three-loop the anomalous dimensions are extracted by imposing the 
universal IR structure of on-shell Form factor in \cite{Ahmed:2016qjf} and 
also collected below for completeness, 
\begin{align}\label{eq:uv-renorm-coefficients}
\gamma_{GG}^{(1)} =& n_f\Big(-\frac{2}{3} \Big)\,, 
\nonumber\\
\gamma_{GG}^{(2)} =& C_A n_f \Big(-\frac{35}{27} \Big) + C_F n_f \Big(-\frac{74}{27}\Big)\,, 
\nonumber\\
\gamma_{GG}^{(3)} =& C_A^2 n_f \Big( - \frac{3589}{162} + 24 \zeta_3\Big)
+ C_A C_F n_f \Big( \frac{139}{9} - \frac{104}{3} \zeta_3 \Big) 
+ C_F^2 n_f \Big(- \frac{2155}{243} + \frac{32}{3} \zeta_3 \Big) 
\nonumber\\
+& C_A n_f^2 \Big( \frac{1058}{243}\Big)
+ C_F n_f^2 \Big(- \frac{173}{243} \Big)\,, 
\nonumber\\
\gamma_{QQ}^{(1)} =& C_F \Big(-\frac{8}{3}\Big)\,,
\nonumber\\
\gamma_{QQ}^{(2)} =& C_A C_F \Big( -\frac{376}{27}\Big) 
+ C_F^2 \Big(\frac{112}{27}\Big) 
+ C_F n_f \Big(\frac{104}{27}\Big)\,,
\nonumber\\
\gamma_{QQ}^{(3)} =& C_A^2 C_F \Big( -\frac{20920}{243} - \frac{64}{3} \zeta_3\Big)
+ C_A C_F^2 \Big( \frac{8528}{243} + 64\zeta_3\Big)
+ C_F^3 \Big(\frac{560}{243} - \frac{128}{3}\zeta_3\Big)
\nonumber\\
+& C_A C_F n_f \Big( \frac{22}{9} + \frac{128}{3} \zeta_3 \Big)
+ C_F^2 n_f \Big(\frac{7094}{243} - \frac{128}{3}\zeta_3 \Big)
+ C_F n_f^2 \Big(\frac{284}{81}\Big)\,.
\end{align}
Note that the conservation of the sum of the operators fixes the 
remaining anomalous dimensions \textit{i.e.}
\begin{align}\label{eq:uv-renorm-connections}
\gamma_{GG} + \gamma_{QG} &= 0 \,,\nonumber\\
\gamma_{GQ} + \gamma_{QQ} &= 0  \,.
\end{align}
\section{Process dependent resum coefficients}\label{app:g0-coefficients}
The process-dependent resum coefficients are collected here for both quark 
and gluon channels with exact dependence of renormalization and 
factorization scales. Notice that the coefficients now explicitly depend 
on the gluon and quark coupling to spin-2 particle through the ratio 
$r_q = k_{g}/k_{q}$ for quark initiated subprocess and through 
$r_g = k_{q}/k_{g}$ for gluon initiated subprocess. At each order 
the coefficients take the following form,
\begin{align}\label{eq:g0-splitted}
 g_{0,p}^{(i),I} = g_{00,p}^{(i),I} + r_p ~g_{01,p}^{(i),I} +  r_p^2 ~g_{02,p}^{(i),I}\,.
\end{align}
Notice that the universal case is recovered after realising $r_p = 1$. 
Defining $\Lqr = \ln(Q^2/\mu_r^2)$ and $\Lfr = \ln(\mu_f^2/\mu_r^2)$,

\begin{align} \label{eq:g0-gg}
\begin{autobreak} 
\gGRGG1 = 
  \grg   \bigg\{ \bigg(\frac{35}{9}
+ \bigg(
- \frac{4}{3}\bigg)  \Lqr\bigg)  \nf \bigg\}      
+ \bigg\{\bigg(
- \frac{203}{9}
+ 16  \z2
+ \bigg(
- \frac{22}{3}\bigg)  \Lfr
+ \bigg(\frac{22}{3}\bigg)  \Lqr\bigg)  \Ca
+ \bigg(\bigg(\frac{4}{3}\bigg)  \Lfr\bigg)  \nf \bigg\}   \,,
\end{autobreak} 
\\ 
\begin{autobreak} 
\gGRGG2 =
  \grg    \bigg\{ \bigg(
- \frac{10295}{162}
+ 16  \z3
+ \frac{442}{9}  \z2
+ \bigg(
- \frac{770}{27}\bigg)  \Lfr
+ \bigg(
- \frac{22}{3}\bigg)  \Lqr^2
+ \bigg(\frac{88}{9}\bigg)  \Lqrfr
+ \bigg(\frac{1127}{27}
- \frac{64}{3}  \z2\bigg)  \Lqr\bigg)  \Ca  \nf
+ \bigg(\frac{598}{81}
- 16  \z3
+ \frac{128}{9}  \z2
+ \bigg(
- \frac{16}{9}\bigg)  \Lqr^2
+ \bigg(\frac{44}{9}\bigg)  \Lqr\bigg)  \Cf  \nf
+ \bigg(\bigg(
- \frac{16}{9}\bigg)   \Lqrfr
+ \bigg(\frac{140}{27}\bigg)  \Lfr\bigg)  \nf^2 \bigg\}      
+ \grg^2    \bigg\{ \bigg(\frac{1225}{324}
+ \frac{8}{3}  \z2
+ \bigg(
- \frac{70}{27}\bigg)  \Lqr
+ \bigg(\frac{4}{9}\bigg)  \Lqr^2\bigg)  \nf^2 \bigg\}      
+ \bigg\{\bigg(\frac{1049}{81}
- \frac{128}{9}  \z2
+ \bigg(
- \frac{80}{9}\bigg)  \Lqr
+ \bigg(\frac{16}{9}\bigg)  \Lqr^2
+ \bigg(4\bigg)  \Lfr\bigg)  \Cf  \nf
+ \bigg(\frac{7801}{324}
+ \frac{88}{9}  \z3
- \frac{1312}{9}  \z2
+ 92  \z2^2
+ \bigg(
- \frac{1657}{27}
+ 24  \z3
+ \frac{176}{3}  \z2\bigg)  \Lqr
+ \bigg(
- \frac{484}{9}\bigg)  \Lqrfr
+ \bigg(\frac{121}{9}\bigg)  \Lqr^2
+ \bigg(\frac{121}{3}\bigg)  \Lfr^2
+ \bigg( \frac{3890}{27}
- 24  \z3
- \frac{352}{3}  \z2\bigg)  \Lfr\bigg)  \Ca^2
+ \bigg(\frac{3656}{81}
- \frac{16}{9}  \z3
- \frac{112}{3}  \z2
+ \bigg(
- \frac{668}{27}
+ \frac{64}{3}  \z2\bigg)  \Lfr
+ \bigg(
- \frac{160}{9}
+ \frac{32}{3}  \z2\bigg)  \Lqr
+ \bigg(
- \frac{44}{3}\bigg)   \Lfr^2
+ \bigg(\frac{22}{9}\bigg)  \Lqr^2
+ \bigg(\frac{88}{9}\bigg)  \Lqrfr\bigg)  \Ca  \nf
+ \bigg(\bigg(\frac{4}{3}\bigg)  \Lfr^2\bigg)  \nf^2 \bigg\}   \,,
\end{autobreak} 
\\ 
\begin{autobreak} 
\gGRGG3 = 
  \grg    \bigg\{ \bigg(
- \frac{464173}{3645}
+ 80  \z5
- \frac{256}{45}  \z3
+ \frac{14708}{81}  \z2
- \frac{704}{3}  \z2   \z3
+ \frac{3376}{15}  \z2^2
+ \bigg(
- \frac{13156}{243}
+ \frac{352}{3}  \z3
- \frac{2816}{27}  \z2\bigg)  \Lfr
+ \bigg(
- \frac{968}{27}\bigg)  \Lqrfr
+ \bigg(
- \frac{176}{27}\bigg)  \Lqr^3
+ \bigg(\frac{352}{27}\bigg)  \Lqrtfr
+ \bigg(\frac{2128}{81}
- \frac{256}{9}  \z2\bigg)  \Lqr^2
+ \bigg(\frac{9530}{243}
- \frac{80}{3}  \z3
+ \frac{1168}{27}  \z2\bigg)  \Lqr\bigg)  \Ca  \Cf  \nf
+ \bigg(
- \frac{2410567}{21870}
- \frac{760}{3}  \z5
+ \frac{50009}{405}  \z3
- \frac{18251}{81}  \z2
+ \frac{716}{3}  \z2   \z3
+ \frac{6449}{45}  \z2^2
+ \bigg(
- \frac{22490}{81}
+ 32  \z3
+ \frac{1408}{9}  \z2\bigg)  \Lqrfr
+ \bigg(
- \frac{484}{9}\bigg)  \Lqrfrt
+ \bigg(
- \frac{484}{81}\bigg)  \Lqr^3
+ \bigg(\frac{5092}{243}
+ \frac{3464}{27}  \z3
+ \frac{1376}{9}  \z2
- \frac{368}{3}  \z2^2\bigg)  \Lqr
+ \bigg(\frac{668}{27}
- 32  \z3
- \frac{352}{9}  \z2\bigg)  \Lqr^2
+ \bigg(\frac{484}{9}\bigg)  \Lqrtfr
+ \bigg(\frac{4235}{27}\bigg)  \Lfr^2
+ \bigg(\frac{93085}{243}
- \frac{632}{3}  \z3
- \frac{9724}{27}  \z2\bigg)  \Lfr\bigg)  \Ca^2  \nf
+ \bigg(
- \frac{35176}{3645}
+ 160  \z5
- \frac{29368}{405}  \z3
- \frac{6016}{81}  \z2
- \frac{64}{15}  \z2^2
+ \bigg(
- \frac{3446}{243}
- \frac{64}{3}  \z3
+ \frac{1024}{27}  \z2\bigg)  \Lqr
+ \bigg(
- \frac{128}{81}\bigg)  \Lqr^3
+ \bigg( \frac{752}{81}\bigg)  \Lqr^2\bigg)  \Cf^2  \nf
+ \bigg(\frac{16087}{486}
+ \frac{3296}{81}  \z3
- \frac{4216}{81}  \z2
+ \bigg(
- \frac{11186}{243}
- \frac{32}{3}  \z3
+ \frac{128}{9}  \z2\bigg)  \Lqr
+ \bigg(
- \frac{64}{27}\bigg)  \Lqrtfr
+ \bigg(
- \frac{160}{81}\bigg)   \Lqr^3
+ \bigg(\frac{32}{27}\bigg)  \Lqrfr
+ \bigg(\frac{1340}{81}\bigg)  \Lqr^2
+ \bigg(\frac{6172}{243}
- \frac{64}{3}  \z3
+ \frac{512}{27}  \z2\bigg)  \Lfr\bigg)  \Cf  \nf^2
+ \bigg(\frac{1404698}{10935}
- \frac{7384}{135}  \z3
- \frac{9112}{81}  \z2
- \frac{32}{45}  \z2^2
+ \bigg(
- \frac{23468}{243}
+ \frac{352}{27}  \z3
+ \frac{2228}{27}  \z2\bigg)  \Lqr
+ \bigg(
- \frac{15550}{243}
+ \frac{64}{3}  \z3
+ \frac{1768}{27}  \z2\bigg)  \Lfr
+ \bigg(
- \frac{1540}{27}\bigg)  \Lfr^2
+ \bigg(
- \frac{88}{9}\bigg)  \Lqrtfr
+ \bigg(
- \frac{220}{81}\bigg)  \Lqr^3
+ \bigg(\frac{176}{9}\bigg)  \Lqrfrt
+ \bigg(\frac{2125}{81}
- \frac{128}{9}  \z2\bigg)  \Lqr^2
+ \bigg(\frac{3932}{81}
- \frac{256}{9}  \z2\bigg)  \Lqrfr\bigg)  \Ca  \nf^2
+ \bigg(\bigg(
- \frac{16}{9}\bigg)  \Lqrfrt
+ \bigg(\frac{140}{27}\bigg)   \Lfr^2\bigg)  \nf^3 \bigg\}      
+ \grg^2    \bigg\{ \bigg(
- \frac{55825}{1458}
+ \frac{280}{9}  \z3
+ \frac{1141}{27}  \z2
+ \frac{128}{3}  \z2^2
+ \bigg(
- \frac{13475}{486}
- \frac{176}{9}  \z2\bigg)  \Lfr
+ \bigg(
- \frac{1057}{81}
+ \frac{64}{9}  \z2\bigg)  \Lqr^2
+ \bigg(
- \frac{88}{27}\bigg)   \Lqrtfr
+ \bigg(\frac{44}{27}\bigg)  \Lqr^3
+ \bigg(\frac{1540}{81}\bigg)  \Lqrfr
+ \bigg(\frac{9070}{243}
- \frac{32}{3}  \z3
- \frac{884}{27}  \z2\bigg)  \Lqr\bigg)  \Ca  \nf^2
+ \bigg(\frac{10465}{729}
- \frac{280}{9}  \z3
+ \frac{656}{81}  \z2
+ \bigg(
- \frac{544}{81}\bigg)   \Lqr^2
+ \bigg(\frac{32}{27}\bigg)  \Lqr^3
+ \bigg(\frac{1114}{243}
+ \frac{32}{3}  \z3
+ \frac{128}{27}  \z2\bigg)  \Lqr\bigg)  \Cf   \nf^2
+ \bigg(\bigg(
- \frac{280}{81}\bigg)  \Lqrfr
+ \bigg(\frac{16}{27}\bigg)  \Lqrtfr
+ \bigg(\frac{1225}{243}
+ \frac{32}{9}  \z2\bigg)  \Lfr \bigg)  \nf^3 \bigg\}      
+\bigg\{ \bigg(
- \frac{303707}{810}
+ \frac{5156}{9}  \z5
- \frac{81074}{135}  \z3
+ 96  \z3^2
- \frac{697}{9}  \z2
+ 48   \z2  \z3
+ \frac{6248}{27}  \z2^2
+ \frac{3872}{15}  \z2^3
+ \bigg(
- \frac{17105}{27}
+ 264  \z3
+ \frac{1936}{3}  \z2\bigg)   \Lfr^2
+ \bigg(
- \frac{5324}{27}\bigg)  \Lfr^3
+ \bigg(
- \frac{1319}{9}
- 160  \z5
- 184  \z3
+ \frac{496}{3}  \z2
+ 352  \z2  \z3
+ \frac{44}{3}  \z2^2\bigg)  \Lqr
+ \bigg(
- \frac{2662}{27}\bigg)  \Lqrtfr
+ \bigg(\frac{110}{3}
+ 88  \z3\bigg)  \Lqr^2
+ \bigg(\frac{109651}{486}
+ 160  \z5
+ \frac{3032}{27}  \z3
+ \frac{19504}{27}  \z2
- 352  \z2  \z3
- \frac{2068}{3}   \z2^2\bigg)  \Lfr
+ \bigg(\frac{23782}{81}
- 352  \z3
- \frac{3872}{9}  \z2\bigg)  \Lqrfr
+ \bigg(\frac{2662}{9}\bigg)  \Lqrfrt\bigg)   \Ca^3
+ \bigg(
- \frac{442583}{4374}
+ \frac{16}{9}  \z3
+ \frac{9196}{81}  \z2
- \frac{688}{45}  \z2^2
+ \bigg(
- \frac{640}{27}
+ \frac{128}{9}  \z2\bigg)  \Lqrfr
+ \bigg(
- \frac{176}{9}\bigg)  \Lfr^3
+ \bigg(
- \frac{464}{27}
+ \frac{64}{3}  \z2\bigg)  \Lfr^2
+ \bigg(
- \frac{320}{27}
+ \frac{64}{9}  \z2\bigg)  \Lqr^2
+ \bigg(\frac{88}{81}\bigg)  \Lqr^3
+ \bigg(\frac{88}{27}\bigg)  \Lqrtfr
+ \bigg(\frac{88}{9}\bigg)  \Lqrfrt
+ \bigg(\frac{13291}{243}
- \frac{64}{27}  \z3
- \frac{448}{9}  \z2\bigg)  \Lqr
+ \bigg(\frac{13841}{243}
- \frac{64}{27}  \z3
- \frac{448}{9}  \z2\bigg)  \Lfr\bigg)  \Ca  \nf^2
+ \bigg(
- \frac{46372}{729}
- \frac{128}{81}  \z3
+ \frac{5504}{81}   \z2
+ \bigg(
- \frac{320}{27}\bigg)  \Lqrfr
+ \bigg(
- \frac{688}{81}\bigg)  \Lqr^2
+ \bigg(\frac{64}{81}\bigg)  \Lqr^3
+ \bigg(\frac{64}{27}\bigg)   \Lqrtfr
+ \bigg(\frac{28}{3}\bigg)  \Lfr^2
+ \bigg(\frac{3602}{243}
- \frac{512}{27}  \z2\bigg)  \Lfr
+ \bigg(\frac{8776}{243}
- \frac{512}{27}  \z2\bigg)  \Lqr\bigg)  \Cf  \nf^2
+ \bigg(
- \frac{62429}{3645}
- \frac{10592}{405}  \z3
+ \frac{6016}{81}  \z2
+ \frac{64}{15}  \z2^2
+ \bigg(
- \frac{752}{81}\bigg)  \Lqr^2
+ \bigg(\frac{128}{81}\bigg)  \Lqr^3
+ \bigg(\frac{3932}{243}
+ \frac{64}{3}  \z3
- \frac{1024}{27}  \z2\bigg)  \Lqr
+ \bigg(-2\bigg)  \Lfr\bigg)  \Cf^2  \nf
+ \bigg(\frac{5410051}{21870}
- \frac{1112}{9}  \z5
+ \frac{95998}{405}  \z3
+ \frac{8804}{81}  \z2
- 144  \z2  \z3
- \frac{42028}{135}  \z2^2
+ \bigg(
- \frac{95572}{243}
+ \frac{2144}{27}  \z3
+ \frac{4592}{27}  \z2
+ \frac{376}{3}  \z2^2\bigg)  \Lfr
+ \bigg(
- \frac{968}{9}\bigg)  \Lqrfrt
+ \bigg(
- \frac{1046}{27}
+ 16  \z3
+ \frac{352}{9}  \z2\bigg)  \Lqr^2
+ \bigg(\frac{484}{81}\bigg)  \Lqr^3
+ \bigg(\frac{9029}{243}
- \frac{2384}{27}  \z3
- \frac{1712}{9}  \z2
+ 120  \z2^2\bigg)  \Lqr
+ \bigg(\frac{7100}{81}
+ 32  \z3\bigg)  \Lqrfr
+ \bigg(\frac{968}{9}\bigg)  \Lfr^3
+ \bigg(\frac{5662}{27}
- 48  \z3
- \frac{704}{3}  \z2\bigg)  \Lfr^2\bigg)  \Ca^2  \nf
+ \bigg( \frac{1056967}{3645}
- \frac{6256}{45}  \z3
- \frac{12152}{81}  \z2
+ 128  \z2  \z3
- \frac{2032}{9}  \z2^2
+ \bigg(
- \frac{38495}{243}
+ \frac{4544}{27}  \z2\bigg)  \Lfr
+ \bigg(
- \frac{154}{3}\bigg)  \Lfr^2
+ \bigg(
- \frac{2722}{81}
+ \frac{256}{9}  \z2\bigg)  \Lqr^2
+ \bigg(
- \frac{5075}{243}
+ \frac{80}{3}  \z3
- \frac{2032}{27}  \z2\bigg)  \Lqr
+ \bigg(
- \frac{352}{27}\bigg)   \Lqrtfr
+ \bigg(\frac{176}{27}\bigg)  \Lqr^3
+ \bigg(\frac{2552}{27}\bigg)  \Lqrfr\bigg)  \Ca  \Cf  \nf
+ \bigg(\bigg(\frac{32}{27}\bigg)  \Lfr^3\bigg)   \nf^3  \bigg\}\,.
\end{autobreak}
\end{align}

\begin{align}\label{eq:g0-qq} 
\begin{autobreak} 
\gGRQQ1 = 
  \grq    \bigg\{ \bigg(\frac{68}{9}
+ \bigg(
- \frac{16}{3}\bigg)  \Lqr\bigg)  \Cf \bigg\}      
+ \bigg\{\bigg(
- \frac{248}{9}
+ 16  \z2
+ \bigg(\frac{34}{3}\bigg)  \Lqr
+ \bigg(-6\bigg)  \Lfr\bigg)  \Cf \bigg\}   \,,
\end{autobreak} 
\\ 
\begin{autobreak} 
\gGRQQ2 = 
  \grq    \bigg\{ \bigg(
- \frac{17998}{81}
+ \frac{832}{9}  \z2
+ \bigg(
- \frac{160}{3}\bigg)  \Lqr^2
+ \bigg(
- \frac{136}{3}\bigg)  \Lfr
+ \bigg(\frac{5960}{27}
- \frac{256}{3}  \z2\bigg)  \Lqr
+ \bigg(32\bigg)  \Lqrfr\bigg)  \Cf^2
+ \bigg(
- \frac{2332}{81}
+ \frac{64}{3}   \z2
+ \bigg(
- \frac{32}{9}\bigg)  \Lqr^2
+ \bigg(\frac{160}{9}\bigg)  \Lqr\bigg)  \Cf  \nf
+ \bigg(\frac{7826}{81}
- \frac{472}{9}  \z2
+ \bigg(
- \frac{500}{9}\bigg)  \Lqr
+ \bigg(\frac{88}{9}\bigg)  \Lqr^2\bigg)  \Ca  \Cf \bigg\}      
+ \grq^2    \bigg\{ \bigg(\frac{1156}{81}
+ \frac{128}{3}  \z2
+ \bigg(
- \frac{544}{27}\bigg)  \Lqr
+ \bigg(\frac{64}{9}\bigg)  \Lqr^2\bigg)  \Cf^2 \bigg\}      
+ \bigg\{\bigg(
- \frac{84773}{324}
+ \frac{1180}{9}  \z3
+ \frac{1336}{9}  \z2
- \frac{92}{5}  \z2^2
+ \bigg(
- \frac{187}{9}\bigg)  \Lqr^2
+ \bigg(
- \frac{17}{3}
+ 24  \z3
- \frac{88}{3}  \z2\bigg)  \Lfr
+ \bigg(\frac{1211}{9}
- 24  \z3
- \frac{88}{3}  \z2\bigg)  \Lqr
+ \bigg( 11\bigg)  \Lfr^2\bigg)  \Ca  \Cf
+ \bigg(\frac{8813}{162}
+ \frac{8}{9}  \z3
- \frac{112}{3}  \z2
+ \bigg(
- \frac{286}{9}
+ \frac{16}{3}  \z2\bigg)   \Lqr
+ \bigg(\frac{2}{3}
+ \frac{16}{3}  \z2\bigg)  \Lfr
+ \bigg(\frac{50}{9}\bigg)  \Lqr^2
+ \bigg(-2\bigg)  \Lfr^2\bigg)  \Cf  \nf
+ \bigg(\frac{43093}{108}
- 124  \z3
- \frac{3286}{9}  \z2
+ \frac{552}{5}  \z2^2
+ \bigg(
- \frac{8575}{27}
+ 48  \z3
+ \frac{472}{3}   \z2\bigg)  \Lqr
+ \bigg(\frac{578}{9}\bigg)  \Lqr^2
+ \bigg(\frac{487}{3}
- 48  \z3
- 72  \z2\bigg)  \Lfr
+ \bigg(-68\bigg)  \Lqrfr
+ \bigg(18\bigg)  \Lfr^2\bigg)  \Cf^2 \bigg\}   \,,
\end{autobreak} 
\\ 
\begin{autobreak} 
\gGRQQ3 = 
  \grq    \bigg\{ \bigg(
- \frac{10514857}{2187}
+ \frac{1280}{3}  \z5
+ \frac{44288}{81}  \z3
+ \frac{270848}{81}  \z2
+ \frac{256}{3}  \z2  \z3
- \frac{15536}{15}  \z2^2
+ \bigg(
- \frac{131140}{81}
+ 128  \z3
+ \frac{2816}{9}  \z2\bigg)   \Lqr^2
+ \bigg(
- \frac{16808}{27}
+ \frac{544}{3}  \z3
+ \frac{2512}{27}  \z2\bigg)  \Lfr
+ \bigg(
- \frac{176}{3}\bigg)  \Lqrfrt
+ \bigg(
- \frac{176}{3}\bigg)  \Lqrtfr
+ \bigg(\frac{748}{9}\bigg)  \Lfr^2
+ \bigg(\frac{1760}{9}\bigg)  \Lqr^3
+ \bigg(\frac{3272}{9}
- 128   \z3
+ \frac{1408}{9}  \z2\bigg)  \Lqrfr
+ \bigg(\frac{1178048}{243}
- \frac{20320}{27}  \z3
- \frac{58000}{27}  \z2
+ \frac{1472}{15}  \z2^2\bigg)  \Lqr\bigg)  \Ca  \Cf^2
+ \bigg(
- \frac{1268944}{2187}
- \frac{10528}{81}  \z3
+ \frac{60512}{81}  \z2
- \frac{896}{45}  \z2^2
+ \bigg(
- \frac{10780}{81}\bigg)  \Lqr^2
+ \bigg(\frac{1232}{81}\bigg)  \Lqr^3
+ \bigg(\frac{107948}{243}
+ \frac{256}{3}  \z3
- \frac{784}{3}  \z2\bigg)  \Lqr\bigg)  \Ca  \Cf  \nf
+ \bigg(\frac{39772}{729}
- \frac{320}{3}  \z2
+ \bigg(
- \frac{4096}{81}
+ \frac{128}{3}  \z2\bigg)  \Lqr
+ \bigg(
- \frac{64}{27}\bigg)  \Lqr^3
+ \bigg(\frac{160}{9}\bigg)  \Lqr^2\bigg)  \Cf   \nf^2
+ \bigg(\frac{2196934}{2187}
+ 160  \z3
- \frac{77536}{81}  \z2
+ \frac{1792}{5}  \z2^2
+ \bigg(
- \frac{280736}{243}
- \frac{2432}{27}  \z3
+ \frac{16064}{27}  \z2\bigg)  \Lqr
+ \bigg(
- \frac{992}{9}
- \frac{256}{9}  \z2\bigg)   \Lqrfr
+ \bigg(
- \frac{4768}{81}\bigg)  \Lqr^3
+ \bigg(
- \frac{136}{9}\bigg)  \Lfr^2
+ \bigg(\frac{32}{3}\bigg)  \Lqrfrt
+ \bigg(\frac{64}{3} \bigg)  \Lqrtfr
+ \bigg(\frac{1600}{9}
- \frac{2368}{27}  \z2\bigg)  \Lfr
+ \bigg(\frac{35560}{81}
- \frac{256}{3}  \z2\bigg)  \Lqr^2\bigg)   \Cf^2  \nf
+ \bigg(\frac{3381256}{2187}
- 480  \z5
+ \frac{29972}{81}  \z3
- \frac{3344}{3}  \z2
- \frac{208}{3}  \z2   \z3
+ \frac{2324}{45}  \z2^2
+ \bigg(
- \frac{234820}{243}
- \frac{128}{3}  \z3
+ \frac{10384}{27}  \z2\bigg)  \Lqr
+ \bigg(
- \frac{1936}{81}\bigg)  \Lqr^3
+ \bigg(\frac{6316}{27}\bigg)  \Lqr^2\bigg)  \Ca^2  \Cf
+ \bigg(\frac{263089}{81}
- \frac{72176}{81}   \z3
- \frac{6184}{3}  \z2
+ \frac{1696}{5}  \z2^2
+ \bigg(
- \frac{1041604}{243}
+ \frac{2816}{3}  \z3
+ \frac{72896}{27}   \z2
- \frac{2944}{5}  \z2^2\bigg)  \Lqr
+ \bigg(
- \frac{11776}{9}
+ 256  \z3
+ 384  \z2\bigg)  \Lqrfr
+ \bigg(
- \frac{21728}{81}\bigg)  \Lqr^3
+ \bigg(\frac{35384}{27}
- \frac{1088}{3}  \z3
- \frac{1120}{3}  \z2\bigg)  \Lfr
+ \bigg(\frac{150136}{81}
- 256  \z3
- \frac{2176}{3}  \z2\bigg)  \Lqr^2
+ \bigg(-96\bigg)  \Lqrfrt
+ \bigg(136\bigg)  \Lfr^2
+ \bigg(320\bigg)   \Lqrtfr\bigg)  \Cf^3 \bigg\}      
+ \grq^2    \bigg\{ \bigg(
- \frac{325244}{729}
- \frac{23360}{27}  \z2
+ \frac{2048}{3}  \z2^2
+ \bigg(
- \frac{29632}{81}
+ \frac{1024}{9}  \z2\bigg)  \Lqr^2
+ \bigg(
- \frac{2312}{27}
- 256  \z2\bigg)  \Lfr
+ \bigg(
- \frac{128}{3}\bigg)  \Lqrtfr
+ \bigg( \frac{1664}{27}\bigg)  \Lqr^3
+ \bigg(\frac{1088}{9}\bigg)  \Lqrfr
+ \bigg(\frac{172408}{243}
+ \frac{256}{27}  \z2\bigg)  \Lqr\bigg)  \Cf^3
+ \bigg(
- \frac{79288}{729}
- \frac{5504}{27}  \z2
+ \bigg(
- \frac{4928}{81}\bigg)  \Lqr^2
+ \bigg(\frac{256}{27}\bigg)  \Lqr^3
+ \bigg(\frac{34976}{243}
+ \frac{512}{9}  \z2\bigg)  \Lqr\bigg)  \Cf^2  \nf
+ \bigg(\frac{266084}{729}
+ \frac{55952}{81}  \z2
+ \bigg(
- \frac{113608}{243}
- \frac{4672}{27}  \z2\bigg)  \Lqr
+ \bigg(
- \frac{704}{27}\bigg)  \Lqr^3
+ \bigg(\frac{14992}{81}\bigg)   \Lqr^2\bigg)  \Ca  \Cf^2 \bigg\}      
+ \bigg\{ \bigg(
- \frac{6131417}{1458}
+ \frac{3344}{3}  \z5
+ \frac{188276}{81}  \z3
+ 32  \z3^2
+ \frac{110552}{27}  \z2
- 1424  \z2  \z3
- \frac{34492}{15}  \z2^2
+ \frac{169504}{315}  \z2^3
+ \bigg(
- \frac{42127}{18}
+ 480  \z5
+ \frac{5792}{3}  \z3
+ \frac{4336}{3}  \z2
- 704  \z2  \z3
- \frac{1968}{5}  \z2^2\bigg)  \Lfr
+ \bigg(
- \frac{49402}{27}
+ 544  \z3
+ \frac{6800}{9}  \z2\bigg)  \Lqr^2
+ \bigg(
- 478
+ 288  \z3
+ 144  \z2\bigg)  \Lfr^2
+ \bigg(
- \frac{1156}{3}\bigg)  \Lqrtfr
+ \bigg(\frac{19652}{81}\bigg)  \Lqr^3
+ \bigg(\frac{16844}{9}
- 832  \z3
- 672  \z2\bigg)  \Lqrfr
+ \bigg(\frac{760175}{162}
- 480  \z5
- \frac{7520}{3}  \z3
- \frac{10576}{3}  \z2
+ 704  \z2  \z3
+ \frac{4912}{5}  \z2^2\bigg)   \Lqr
+ \bigg(-36\bigg)  \Lfr^3
+ \bigg(204\bigg)  \Lqrfrt\bigg)  \Cf^3
+ \bigg(
- \frac{31255393}{8748}
- \frac{2852}{3}  \z5
+ \frac{309353}{81}  \z3
- \frac{400}{3}  \z3^2
+ \frac{181355}{81}  \z2
- \frac{7036}{9}  \z2  \z3
- \frac{3529}{27}   \z2^2
+ \frac{7088}{63}  \z2^3
+ \bigg(
- \frac{15055}{27}
+ 88  \z3
+ \frac{968}{9}  \z2\bigg)  \Lqr^2
+ \bigg(
- \frac{242}{9}\bigg)  \Lfr^3
+ \bigg(\frac{4114}{81}\bigg)  \Lqr^3
+ \bigg(\frac{493}{9}
- 88  \z3
+ \frac{968}{9}  \z2\bigg)  \Lfr^2
+ \bigg( \frac{1657}{18}
- 80  \z5
+ \frac{3104}{9}  \z3
- \frac{8992}{27}  \z2
+ 4  \z2^2\bigg)  \Lfr
+ \bigg(\frac{561610}{243}
+ 80  \z5
- \frac{34120}{27}  \z3
- \frac{8432}{9}  \z2
+ \frac{1964}{15}  \z2^2\bigg)  \Lqr\bigg)  \Ca^2  \Cf
+ \bigg(
- \frac{2467183}{2187}
- \frac{608}{9}  \z5
+ \frac{13960}{27}  \z3
+ \frac{115336}{81}  \z2
- \frac{256}{3}  \z2  \z3
- \frac{75568}{135}  \z2^2
+ \bigg(
- \frac{40028}{81}
+ 32  \z3
+ \frac{400}{3}  \z2\bigg)  \Lqr^2
+ \bigg(
- \frac{2689}{9}
+ \frac{256}{3}  \z3
+ \frac{2008}{27}  \z2
+ \frac{272}{5}  \z2^2\bigg)  \Lfr
+ \bigg(
- \frac{100}{3}\bigg)  \Lqrtfr
+ \bigg(
- \frac{68}{3}\bigg)  \Lqrfrt
+ \bigg(\frac{388}{9}
- 32  \z3
- 48  \z2\bigg)  \Lfr^2
+ \bigg(\frac{4972}{81}\bigg)  \Lqr^3
+ \bigg( \frac{1784}{9}
+ \frac{256}{9}  \z2\bigg)  \Lqrfr
+ \bigg(\frac{109118}{81}
- \frac{4336}{27}  \z3
- \frac{24656}{27}  \z2
+ \frac{464}{5}  \z2^2\bigg)  \Lqr
+ \bigg(12\bigg)  \Lfr^3\bigg)  \Cf^2  \nf
+ \bigg(
- \frac{61807}{729}
- \frac{1136}{81}  \z3
+ \frac{3728}{27}  \z2
+ \frac{448}{135}  \z2^2
+ \bigg(
- \frac{244}{9}
+ \frac{32}{9}  \z2\bigg)  \Lqr^2
+ \bigg(
- \frac{8}{9}\bigg)   \Lfr^3
+ \bigg(\frac{4}{9}
+ \frac{32}{9}  \z2\bigg)  \Lfr^2
+ \bigg(\frac{88}{27}\bigg)  \Lqr^3
+ \bigg(\frac{34}{9}
+ \frac{32}{9}  \z3
- \frac{160}{27}  \z2\bigg)  \Lfr
+ \bigg(\frac{6556}{81}
- \frac{64}{27}  \z3
- \frac{1568}{27}  \z2\bigg)  \Lqr\bigg)  \Cf  \nf^2
+ \bigg( \frac{2446783}{2187}
+ \frac{136}{3}  \z5
- \frac{37456}{81}  \z3
- \frac{93268}{81}  \z2
+ \frac{928}{9}  \z2  \z3
+ \frac{1612}{135}  \z2^2
+ \bigg(
- \frac{213536}{243}
+ \frac{3440}{27}  \z3
+ \frac{12448}{27}  \z2
- \frac{344}{15}   \z2^2\bigg)  \Lqr
+ \bigg(
- 40
- \frac{400}{9}  \z3
+ \frac{2672}{27}  \z2
- \frac{8}{5}  \z2^2\bigg)  \Lfr
+ \bigg(
- \frac{2024}{81}\bigg)  \Lqr^3
+ \bigg(
- \frac{146}{9}
+ 16  \z3
- \frac{352}{9}  \z2\bigg)  \Lfr^2
+ \bigg(\frac{88}{9}\bigg)  \Lfr^3
+ \bigg( \frac{20014}{81}
- 16  \z3
- \frac{352}{9}  \z2\bigg)  \Lqr^2\bigg)  \Ca  \Cf  \nf
+ \bigg(\frac{65475811}{8748}
+ \frac{3128}{9}  \z5
- \frac{546572}{81}  \z3
+ \frac{592}{3}  \z3^2
- \frac{503536}{81}  \z2
+ 2208  \z2  \z3
+ \frac{350464}{135}  \z2^2
- \frac{123632}{315}  \z2^3
+ \bigg(
- \frac{3309509}{486}
+ 240  \z5
+ \frac{94264}{27}  \z3
+ \frac{102992}{27}  \z2
- 352  \z2  \z3
- \frac{10208}{15}  \z2^2\bigg)  \Lqr
+ \bigg(
- \frac{7844}{9}
+ 416   \z3
- \frac{1408}{9}  \z2\bigg)  \Lqrfr
+ \bigg(
- \frac{2323}{9}
+ 32  \z3
+ 264  \z2\bigg)  \Lfr^2
+ \bigg(
- \frac{6358}{27}\bigg)  \Lqr^3
+ \bigg(\frac{374}{3}\bigg)  \Lqrfrt
+ \bigg(\frac{374}{3}\bigg)  \Lqrtfr
+ \bigg(\frac{44564}{27}
- 240  \z5
- \frac{6032}{3}  \z3
+ \frac{2708}{27}  \z2
+ 352  \z2  \z3
- \frac{1136}{5}  \z2^2\bigg)  \Lfr
+ \bigg(\frac{57157}{27}
- 448  \z3
- \frac{5192}{9}  \z2\bigg)  \Lqr^2
+ \bigg(-66\bigg)  \Lfr^3\bigg)  \Ca  \Cf^2 \bigg\} \,.
\end{autobreak} 
\end{align}

Note that all the coefficients for the SM DY process up to third order 
have been obtained by some of us previously in \cite{H.:2020ecd}. 

\section{Soft-Virtual coefficients at the third order}\label{app:sv-coefficints}
Here we collect the new third order soft-virtual coefficients for both 
gluon and quark channels.
\begin{align}\label{eq:svcoeff-gg} 
\begin{autobreak}
\CIg3 =  
 \Dm1     {\color{MidnightBlue} \bigg\{ }  \bigg(
- \frac{303707}{810}
- \frac{28636}{9}  \z5
- \frac{186194}{135}  \z3
+ \frac{13216}{3}  \z3^2
+ \frac{923}{3}  \z2
+ \frac{8944}{3}  \z2  \z3
+ \frac{29416}{135}  \z2^2
- \frac{64096}{105}  \z2^3
+ \bigg(
- \frac{1319}{9}
+ 5984  \z5
- \frac{3496}{3}  \z3
+ \frac{15232}{27}  \z2
- 3872  \z2  \z3
- \frac{396}{5}  \z2^2\bigg)   \Lqf
+ \bigg(\frac{110}{3}
+ \frac{968}{3}  \z3
+ \frac{736}{3}  \z2
- \frac{1152}{5}  \z2^2\bigg)  \Lqf^2
+ \bigg(\frac{512}{3}  \z3
- \frac{352}{3}  \z2\bigg)  \Lqf^3\bigg)  \Ca^3
+ \bigg(
- \frac{442583}{4374}
- \frac{1136}{81}  \z3
+ \frac{8396}{81}  \z2
- \frac{208}{15}  \z2^2
+ \bigg(
- \frac{320}{27}
+ \frac{32}{9}  \z2\bigg)  \Lqf^2
+ \bigg(\frac{88}{81}\bigg)  \Lqf^3
+ \bigg(\frac{13291}{243}
+ \frac{64}{9}  \z3
- \frac{1024}{27}  \z2 \bigg)  \Lqf\bigg)  \Ca  \nf^2
+ \bigg(
- \frac{46372}{729}
- \frac{128}{81}  \z3
+ \frac{5504}{81}  \z2
+ \bigg(
- \frac{688}{81}\bigg)  \Lqf^2
+ \bigg(\frac{64}{81}\bigg)   \Lqf^3
+ \bigg(\frac{8776}{243}
- \frac{512}{27}  \z2\bigg)  \Lqf\bigg)  \Cf  \nf^2
+ \bigg(
- \frac{62429}{3645}
- \frac{10592}{405}  \z3
+ \frac{6016}{81}  \z2
+ \frac{64}{15}  \z2^2
+ \bigg(
- \frac{752}{81}\bigg)   \Lqf^2
+ \bigg(\frac{128}{81}\bigg)  \Lqf^3
+ \bigg(\frac{3932}{243}
+ \frac{64}{3}  \z3
- \frac{1024}{27}  \z2\bigg)  \Lqf\bigg)  \Cf^2   \nf
+ \bigg(\frac{5410051}{21870}
+ \frac{5032}{9}  \z5
+ \frac{177278}{405}  \z3
- \frac{20276}{81}  \z2
- \frac{1840}{3}  \z2   \z3
+ \frac{964}{27}  \z2^2
+ \bigg(
- \frac{1046}{27}
+ 144  \z3
+ \frac{272}{3}  \z2\bigg)  \Lqf^2
+ \bigg(\frac{484}{81}
- \frac{64}{3}  \z2\bigg)  \Lqf^3
+ \bigg(\frac{9029}{243}
- \frac{3824}{9}  \z3
- \frac{2704}{27}  \z2
+ \frac{40}{3}  \z2^2\bigg)   \Lqf\bigg)  \Ca^2  \nf
+ \bigg(\frac{1056967}{3645}
- \frac{5296}{45}  \z3
- \frac{4868}{27}  \z2
+ 64  \z2  \z3
- 112  \z2^2
+ \bigg(
- \frac{2722}{81}
+ \frac{128}{9}  \z2\bigg)  \Lqf^2
+ \bigg(
- \frac{5075}{243}
+ \frac{80}{3}  \z3
- \frac{544}{27}  \z2\bigg)  \Lqf
+ \bigg(\frac{176}{27}\bigg)  \Lqf^3\bigg)  \Ca  \Cf  \nf
+ \bigg(\bigg(
- \frac{464173}{3645}
+ 80  \z5
- \frac{256}{45}  \z3
+ \frac{3308}{27}  \z2
- \frac{320}{3}  \z2  \z3
+ \frac{5008}{45}  \z2^2
+ \bigg(
- \frac{176}{27}\bigg)  \Lqf^3
+ \bigg(\frac{2128}{81}
- \frac{128}{9}  \z2\bigg)  \Lqf^2
+ \bigg(\frac{9530}{243}
- \frac{80}{3}  \z3
+ \frac{112}{27}  \z2\bigg)  \Lqf\bigg)  \Ca  \Cf  \nf
+ \bigg(
- \frac{2410567}{21870}
- \frac{760}{3}   \z5
- \frac{11591}{405}  \z3
+ \frac{4169}{81}  \z2
+ \frac{332}{3}  \z2  \z3
- \frac{5071}{45}  \z2^2
+ \bigg(
- \frac{484}{81}
+ \frac{128}{3}  \z2\bigg)  \Lqf^3
+ \bigg(\frac{5092}{243}
+ \frac{6104}{9}  \z3
+ \frac{112}{9}  \z2
+ \frac{16}{15}  \z2^2\bigg)   \Lqf
+ \bigg(\frac{668}{27}
- \frac{608}{3}  \z3
- 144  \z2\bigg)  \Lqf^2\bigg)  \Ca^2  \nf
+ \bigg(
- \frac{35176}{3645}
+ 160  \z5
- \frac{29368}{405}  \z3
- \frac{6016}{81}  \z2
- \frac{64}{15}  \z2^2
+ \bigg(
- \frac{3446}{243}
- \frac{64}{3}  \z3
+ \frac{1024}{27}  \z2\bigg)  \Lqf
+ \bigg(
- \frac{128}{81}\bigg)  \Lqf^3
+ \bigg(\frac{752}{81}\bigg)  \Lqf^2\bigg)  \Cf^2   \nf
+ \bigg(\frac{16087}{486}
+ \frac{3296}{81}  \z3
- \frac{4216}{81}  \z2
+ \bigg(
- \frac{11186}{243}
- \frac{32}{3}  \z3
+ \frac{128}{9}  \z2\bigg)  \Lqf
+ \bigg(
- \frac{160}{81}\bigg)  \Lqf^3
+ \bigg(\frac{1340}{81}\bigg)  \Lqf^2\bigg)  \Cf  \nf^2
+ \bigg( \frac{1404698}{10935}
- \frac{10952}{405}  \z3
- \frac{2104}{27}  \z2
- \frac{32}{45}  \z2^2
+ \bigg(
- \frac{23468}{243}
+ \frac{32}{9}  \z3
+ \frac{1348}{27}  \z2\bigg)  \Lqf
+ \bigg(
- \frac{220}{81}\bigg)  \Lqf^3
+ \bigg(\frac{2125}{81}
- \frac{64}{9}   \z2\bigg)  \Lqf^2\bigg)  \Ca  \nf^2\bigg)  \rqg
+ \bigg(\bigg(
- \frac{55825}{1458}
+ \frac{280}{9}  \z3
+ \frac{973}{81}  \z2
+ \frac{64}{3}  \z2^2
+ \bigg(
- \frac{1057}{81}
+ \frac{32}{9}  \z2\bigg)  \Lqf^2
+ \bigg(\frac{44}{27}\bigg)  \Lqf^3
+ \bigg(\frac{9070}{243}
- \frac{32}{3}  \z3
- 12  \z2\bigg)  \Lqf\bigg)  \Ca   \nf^2
+ \bigg(\frac{10465}{729}
- \frac{280}{9}  \z3
+ \frac{656}{81}  \z2
+ \bigg(
- \frac{544}{81}\bigg)  \Lqf^2
+ \bigg(\frac{32}{27} \bigg)  \Lqf^3
+ \bigg(\frac{1114}{243}
+ \frac{32}{3}  \z3
+ \frac{128}{27}  \z2\bigg)  \Lqf\bigg)  \Cf  \nf^2\bigg)  \rqg^2  {\color{MidnightBlue} \bigg\} }      
+ \D5     {\color{MidnightBlue} \bigg\{ }  \bigg(512\bigg)  \Ca^3  {\color{MidnightBlue} \bigg\} }      
+ \D4     {\color{MidnightBlue} \bigg\{ }  \bigg(
- \frac{7040}{9}
+ \bigg(1280\bigg)  \Lqf\bigg)  \Ca^3
+ \bigg(\frac{1280}{9}\bigg)  \Ca^2  \nf  {\color{MidnightBlue} \bigg\} }      
+ \D3     {\color{MidnightBlue} \bigg\{ }  \bigg(
- \frac{18752}{27}
- 3584  \z2
+ \bigg(
- \frac{5632}{9}\bigg)  \Lqf
+ \bigg(1024\bigg)  \Lqf^2\bigg)  \Ca^3
+ \bigg(
- \frac{10496}{27}
+ \bigg(\frac{2560}{9}\bigg)  \Lqf\bigg)  \Ca^2  \nf
+ \bigg(\frac{256}{27}\bigg)  \Ca  \nf^2
+ \bigg(\bigg(\frac{4480}{9}
+ \bigg(
- \frac{512}{3}\bigg)  \Lqf\bigg)  \Ca^2  \nf\bigg)  \rqg  {\color{MidnightBlue} \bigg\} }      
+ \D2     {\color{MidnightBlue} \bigg\{ }  \bigg(
- 1168
+ 11584  \z3
+ \frac{11968}{3}  \z2
+ \bigg(
- 1472
- 5376  \z2\bigg)  \Lqf
+ \bigg( 256\bigg)  \Lqf^3
+ \bigg(352\bigg)  \Lqf^2\bigg)  \Ca^3
+ \bigg(
- \frac{640}{27}
+ \bigg(\frac{128}{9}\bigg)  \Lqf\bigg)  \Ca  \nf^2
+ \bigg(\frac{8128}{27}
- \frac{2176}{3}  \z2
+ \bigg(
- \frac{4544}{9}\bigg)  \Lqf
+ \bigg(192\bigg)  \Lqf^2\bigg)  \Ca^2  \nf
+ \bigg(\bigg(
- \frac{6160}{27}
+ \bigg(\frac{7424}{9}\bigg)  \Lqf
+ \bigg(-256\bigg)  \Lqf^2\bigg)  \Ca^2  \nf
+ \bigg(\frac{1120}{27}
+ \bigg(
- \frac{128}{9}\bigg)  \Lqf\bigg)  \Ca  \nf^2\bigg)  \rqg
+ \bigg(32\bigg)  \Ca  \Cf  \nf  {\color{MidnightBlue} \bigg\} }      
+ \D1     {\color{MidnightBlue} \bigg\{ }  \bigg(
- \frac{6932}{9}
- \frac{20416}{3}  \z3
+ \frac{17120}{9}  \z2
- \frac{9856}{5}  \z2^2
+ \bigg(
- \frac{21536}{27}
+ 11520  \z3
+ \frac{5632}{3}  \z2\bigg)  \Lqf
+ \bigg(
- \frac{1472}{3}
- 2304  \z2\bigg)  \Lqf^2
+ \bigg(\frac{704}{3} \bigg)  \Lqf^3\bigg)  \Ca^3
+ \bigg(\frac{1600}{81}
- \frac{256}{9}  \z2
+ \bigg(
- \frac{640}{27}\bigg)  \Lqf
+ \bigg(\frac{64}{9}\bigg)  \Lqf^2\bigg)  \Ca  \nf^2
+ \bigg(\frac{4904}{81}
+ 128  \z3
- \frac{2048}{9}  \z2
+ \bigg(
- \frac{992}{9}\bigg)  \Lqf
+ \bigg(\frac{256}{9}\bigg)  \Lqf^2\bigg)  \Ca  \Cf   \nf
+ \bigg(\frac{58160}{81}
+ \frac{3328}{3}  \z3
+ \frac{3200}{9}  \z2
+ \bigg(
- \frac{4864}{27}
- \frac{1664}{3}  \z2\bigg)  \Lqf
+ \bigg(
- \frac{928}{9}\bigg)  \Lqf^2
+ \bigg(\frac{128}{3}\bigg)  \Lqf^3\bigg)  \Ca^2  \nf
+ \bigg(\bigg(
- \frac{44840}{81}
+ 256  \z3
- 832  \z2
+ \bigg(
- \frac{256}{3}\bigg)  \Lqf^3
+ \bigg(\frac{1888}{9}\bigg)  \Lqf^2
+ \bigg(\frac{2528}{9}
+ \frac{640}{3}  \z2\bigg)  \Lqf\bigg)  \Ca^2  \nf
+ \bigg(
- \frac{5600}{81}
+ \bigg(
- \frac{128}{9}\bigg)  \Lqf^2
+ \bigg(\frac{1760}{27}\bigg)  \Lqf\bigg)  \Ca  \nf^2
+ \bigg(\frac{9568}{81}
- 256  \z3
+ \frac{2048}{9}  \z2
+ \bigg(
- \frac{256}{9} \bigg)  \Lqf^2
+ \bigg(\frac{704}{9}\bigg)  \Lqf\bigg)  \Ca  \Cf  \nf\bigg)  \rqg
+ \bigg(\bigg(\frac{4900}{81}
+ \frac{128}{3}  \z2
+ \bigg(
- \frac{1120}{27}\bigg)  \Lqf
+ \bigg(\frac{64}{9}\bigg)  \Lqf^2\bigg)  \Ca  \nf^2\bigg)   \rqg^2  {\color{MidnightBlue} \bigg\} }      
+ \D0     {\color{MidnightBlue} \bigg\{ }  \bigg(
- \frac{11164}{729}
- \frac{7600}{9}  \z3
- \frac{15280}{81}  \z2
- \frac{544}{15}  \z2^2
+ \bigg(
- \frac{4912}{27}
- \frac{224}{3}  \z2\bigg)  \Lqf^2
+ \bigg(\frac{704}{27}\bigg)  \Lqf^3
+ \bigg(\frac{11336}{27}
+ \frac{1664}{3}  \z3
+ \frac{896}{9}  \z2\bigg)  \Lqf\bigg)  \Ca^2  \nf
+ \bigg(
- \frac{3712}{729}
+ \frac{320}{27}  \z3
+ \frac{640}{27}  \z2
+ \bigg(
- \frac{160}{27}\bigg)  \Lqf^2
+ \bigg(\frac{32}{27}\bigg)   \Lqf^3
+ \bigg(\frac{800}{81}
- \frac{128}{9}  \z2\bigg)  \Lqf\bigg)  \Ca  \nf^2
+ \bigg(\frac{3422}{27}
- \frac{608}{9}  \z3
- 32  \z2
- \frac{64}{5}  \z2^2
+ \bigg(
- \frac{568}{9}\bigg)  \Lqf^2
+ \bigg(\frac{128}{9}\bigg)   \Lqf^3
+ \bigg(\frac{2452}{81}
+ 64  \z3
- \frac{1024}{9}  \z2\bigg)  \Lqf\bigg)  \Ca  \Cf  \nf
+ \bigg(\frac{390086}{729}
+ 11904  \z5
- \frac{46952}{27}  \z3
+ \frac{29824}{81}  \z2
- \frac{23200}{3}  \z2  \z3
+ \frac{4048}{15}  \z2^2
+ \bigg(
- \frac{66746}{81}
- \frac{3344}{3}  \z3
+ \frac{4144}{3}  \z2
- \frac{4928}{5}  \z2^2\bigg)   \Lqf
+ \bigg(\frac{116}{9}
+ 2240  \z3
- 176  \z2\bigg)  \Lqf^2
+ \bigg(\frac{968}{27}
- 256  \z2\bigg)  \Lqf^3\bigg)   \Ca^3
+ \bigg(\bigg(
- \frac{56560}{243}
+ \frac{3640}{3}  \z3
+ \frac{6160}{27}  \z2
+ \bigg(
- \frac{15956}{81}
- 288  \z3
- \frac{4448}{9}  \z2\bigg)  \Lqf
+ \bigg(
- \frac{352}{9}\bigg)  \Lqf^3
+ \bigg(\frac{5332}{27}
+ \frac{320}{3}  \z2\bigg)  \Lqf^2\bigg)  \Ca^2   \nf
+ \bigg(\frac{7840}{243}
- \frac{1120}{27}  \z2
+ \bigg(
- \frac{1232}{27}
+ \frac{128}{9}  \z2\bigg)  \Lqf
+ \bigg(
- \frac{32}{9}\bigg)   \Lqf^3
+ \bigg(\frac{200}{9}\bigg)  \Lqf^2\bigg)  \Ca  \nf^2
+ \bigg(\bigg(
- \frac{128}{9}\bigg)  \Lqf^3
+ \bigg(\frac{352}{9}\bigg)  \Lqf^2
+ \bigg(\frac{4784}{81}
- 128  \z3
+ \frac{1024}{9}  \z2\bigg)  \Lqf\bigg)  \Ca  \Cf  \nf\bigg)  \rqg
+ \bigg(\bigg(\bigg(
- \frac{560}{27}\bigg)  \Lqf^2
+ \bigg(\frac{32}{9}\bigg)  \Lqf^3
+ \bigg(\frac{2450}{81}
+ \frac{64}{3}  \z2\bigg)  \Lqf\bigg)  \Ca  \nf^2 \bigg)  \rqg^2  {\color{MidnightBlue} \bigg\} } 
\end{autobreak}
\end{align}

\begin{align}\label{eq:svcoeff-qqb} 
\begin{autobreak}
\CIq3 = 
 \Dm1     {\color{MidnightBlue} \bigg\{ }  \bigg(
- \frac{6131417}{1458}
+ \frac{3344}{3}  \z5
+ \frac{188276}{81}  \z3
+ \frac{10336}{3}  \z3^2
+ \frac{8122}{9}  \z2
- 432  \z2  \z3
+ \frac{4156}{45}  \z2^2
- \frac{184736}{315}  \z2^3
+ \bigg(
- \frac{49402}{27}
+ \frac{5984}{3}  \z3
+ \frac{10112}{9}  \z2
- \frac{1792}{5}  \z2^2\bigg)  \Lqf^2
+ \bigg(\frac{19652}{81}
+ \frac{512}{3}   \z3
- \frac{1088}{3}  \z2\bigg)  \Lqf^3
+ \bigg(\frac{760175}{162}
+ 5664  \z5
- \frac{54304}{9}  \z3
- \frac{26584}{27}   \z2
- 2752  \z2  \z3
- \frac{176}{3}  \z2^2\bigg)  \Lqf\bigg)  \Cf^3
+ \bigg(
- \frac{31255393}{8748}
- \frac{2852}{3}  \z5
+ \frac{84131}{27}  \z3
- \frac{400}{3}  \z3^2
+ \frac{39781}{27}   \z2
- \frac{820}{3}  \z2  \z3
+ \frac{20323}{135}  \z2^2
+ \frac{13264}{315}  \z2^3
+ \bigg(
- \frac{15055}{27}
+ 88  \z3\bigg)  \Lqf^2
+ \bigg(\frac{4114}{81}\bigg)  \Lqf^3
+ \bigg(\frac{561610}{243}
+ 80  \z5
- \frac{8792}{9}  \z3
- \frac{11056}{27}  \z2
+ \frac{68}{5}  \z2^2\bigg)  \Lqf\bigg)  \Ca^2  \Cf
+ \bigg(
- \frac{2467183}{2187}
+ \frac{5536}{9}  \z5
+ \frac{38488}{81}  \z3
+ \frac{66184}{81}  \z2
- \frac{5504}{9}  \z2   \z3
- \frac{21808}{135}  \z2^2
+ \bigg(
- \frac{40028}{81}
+ 160  \z3
+ \frac{896}{9}  \z2\bigg)  \Lqf^2
+ \bigg(\frac{4972}{81}
- \frac{64}{3}  \z2\bigg)  \Lqf^3
+ \bigg(\frac{109118}{81}
- \frac{3280}{9}  \z3
- \frac{13328}{27}  \z2
+ \frac{112}{15}   \z2^2\bigg)  \Lqf\bigg)  \Cf^2  \nf
+ \bigg(
- \frac{61807}{729}
- \frac{2416}{81}  \z3
+ \frac{10384}{81}  \z2
+ \frac{128}{27}  \z2^2
+ \bigg(
- \frac{244}{9}\bigg)   \Lqf^2
+ \bigg(\frac{88}{27}\bigg)  \Lqf^3
+ \bigg(\frac{6556}{81}
+ \frac{64}{9}  \z3
- \frac{416}{9}  \z2\bigg)  \Lqf\bigg)  \Cf  \nf^2
+ \bigg(\frac{2446783}{2187}
+ \frac{136}{3}  \z5
- \frac{6320}{27}  \z3
- \frac{8540}{9}  \z2
+ \frac{224}{3}  \z2  \z3
- \frac{1060}{27}  \z2^2
+ \bigg(
- \frac{213536}{243}
+ \frac{208}{9}  \z3
+ \frac{2608}{9}  \z2
- \frac{8}{5}  \z2^2\bigg)  \Lqf
+ \bigg(
- \frac{2024}{81}\bigg)  \Lqf^3
+ \bigg(\frac{20014}{81}
- 16  \z3\bigg)  \Lqf^2\bigg)  \Ca  \Cf  \nf
+ \bigg(\frac{65475811}{8748}
- \frac{30664}{9}  \z5
- \frac{178948}{27}  \z3
+ \frac{3280}{3}  \z3^2
- \frac{201062}{81}   \z2
+ \frac{32960}{9}  \z2  \z3
+ \frac{40576}{135}  \z2^2
- \frac{20816}{315}  \z2^3
+ \bigg(
- \frac{3309509}{486}
+ 240  \z5
+ \frac{44584}{9}  \z3
+ \frac{46808}{27}  \z2
- 1120  \z2  \z3
- \frac{448}{15}  \z2^2\bigg)   \Lqf
+ \bigg(
- \frac{6358}{27}
+ \frac{352}{3}  \z2\bigg)  \Lqf^3
+ \bigg(\frac{57157}{27}
- 1152  \z3
- \frac{1664}{3}  \z2
+ 128  \z2^2\bigg)  \Lqf^2\bigg)  \Ca  \Cf^2
+ \bigg(\bigg(
- \frac{10514857}{2187}
+ \frac{1280}{3}  \z5
+ \frac{6784}{27}  \z3
+ \frac{19088}{9}  \z2
+ \frac{256}{3}  \z2   \z3
- \frac{22288}{45}  \z2^2
+ \bigg(
- \frac{131140}{81}
+ 128  \z3
+ \frac{704}{9}  \z2\bigg)  \Lqf^2
+ \bigg(\frac{1760}{9}\bigg)  \Lqf^3
+ \bigg(\frac{1178048}{243}
- 544  \z3
- \frac{10480}{9}  \z2
+ \frac{64}{5}  \z2^2\bigg)  \Lqf\bigg)  \Ca   \Cf^2
+ \bigg(
- \frac{1268944}{2187}
- \frac{10528}{81}  \z3
+ \frac{60512}{81}  \z2
- \frac{896}{45}  \z2^2
+ \bigg(
- \frac{10780}{81}\bigg)  \Lqf^2
+ \bigg(\frac{1232}{81}\bigg)  \Lqf^3
+ \bigg(\frac{107948}{243}
+ \frac{256}{3}  \z3
- \frac{784}{3}   \z2\bigg)  \Lqf\bigg)  \Ca  \Cf  \nf
+ \bigg(\frac{39772}{729}
- \frac{320}{3}  \z2
+ \bigg(
- \frac{4096}{81}
+ \frac{128}{3}  \z2\bigg)   \Lqf
+ \bigg(
- \frac{64}{27}\bigg)  \Lqf^3
+ \bigg(\frac{160}{9}\bigg)  \Lqf^2\bigg)  \Cf  \nf^2
+ \bigg(\frac{2196934}{2187}
+ \frac{17312}{81}  \z3
- \frac{53440}{81}  \z2
+ \frac{2816}{15}  \z2^2
+ \bigg(
- \frac{280736}{243}
- 128  \z3
+ \frac{9856}{27}  \z2\bigg)  \Lqf
+ \bigg(
- \frac{4768}{81}\bigg)  \Lqf^3
+ \bigg(\frac{35560}{81}
- \frac{256}{9}  \z2\bigg)  \Lqf^2\bigg)   \Cf^2  \nf
+ \bigg(\frac{3381256}{2187}
- 480  \z5
+ \frac{29972}{81}  \z3
- \frac{3344}{3}  \z2
- \frac{208}{3}  \z2   \z3
+ \frac{2324}{45}  \z2^2
+ \bigg(
- \frac{234820}{243}
- \frac{128}{3}  \z3
+ \frac{10384}{27}  \z2\bigg)  \Lqf
+ \bigg(
- \frac{1936}{81}\bigg)  \Lqf^3
+ \bigg(\frac{6316}{27}\bigg)  \Lqf^2\bigg)  \Ca^2  \Cf
+ \bigg(\frac{263089}{81}
- \frac{72176}{81}   \z3
- \frac{22984}{81}  \z2
- \frac{11488}{45}  \z2^2
+ \bigg(
- \frac{1041604}{243}
+ \frac{17152}{9}  \z3
+ \frac{25216}{27}  \z2
- \frac{128}{15}  \z2^2\bigg)  \Lqf
+ \bigg(
- \frac{21728}{81}
+ \frac{512}{3}  \z2\bigg)  \Lqf^3
+ \bigg( \frac{150136}{81}
- \frac{2816}{3}  \z3
- \frac{4864}{9}  \z2\bigg)  \Lqf^2\bigg)  \Cf^3\bigg)  \rgq
+ \bigg(\bigg(
- \frac{325244}{729}
- \frac{79328}{81}  \z2
+ \frac{1024}{3}  \z2^2
+ \bigg(
- \frac{29632}{81}
+ \frac{512}{9}  \z2 \bigg)  \Lqf^2
+ \bigg(\frac{1664}{27}\bigg)  \Lqf^3
+ \bigg(\frac{172408}{243}
+ \frac{512}{3}  \z2\bigg)  \Lqf\bigg)  \Cf^3
+ \bigg(
- \frac{79288}{729}
- \frac{5504}{27}  \z2
+ \bigg(
- \frac{4928}{81}\bigg)  \Lqf^2
+ \bigg(\frac{256}{27}\bigg)  \Lqf^3
+ \bigg(\frac{34976}{243}
+ \frac{512}{9}  \z2\bigg)  \Lqf\bigg)  \Cf^2  \nf
+ \bigg(\frac{266084}{729}
+ \frac{55952}{81}  \z2
+ \bigg(
- \frac{113608}{243}
- \frac{4672}{27}  \z2\bigg)  \Lqf
+ \bigg(
- \frac{704}{27}\bigg)  \Lqf^3
+ \bigg(\frac{14992}{81}\bigg)  \Lqf^2\bigg)  \Ca  \Cf^2 \bigg)  \rgq^2  {\color{MidnightBlue} \bigg\} }      
+ \D5     {\color{MidnightBlue} \bigg\{ }  \bigg(512\bigg)  \Cf^3  {\color{MidnightBlue} \bigg\} }      
+ \D4     {\color{MidnightBlue} \bigg\{ }  \bigg(
- \frac{7040}{9}\bigg)  \Ca  \Cf^2
+ \bigg(\frac{1280}{9}\bigg)  \Cf^2  \nf
+ \bigg(\bigg(1280\bigg)  \Lqf\bigg)  \Cf^3  {\color{MidnightBlue} \bigg\} }      
+ \D3     {\color{MidnightBlue} \bigg\{ }  \bigg(
- \frac{31744}{9}
- 3072  \z2
+ \bigg(\frac{4352}{3}\bigg)  \Lqf
+ \bigg(1024\bigg)  \Lqf^2\bigg)  \Cf^3
+ \bigg(
- \frac{2560}{9}
+ \bigg(\frac{2560}{9}\bigg)  \Lqf\bigg)  \Cf^2  \nf
+ \bigg(
- \frac{2816}{27}\bigg)  \Ca  \Cf  \nf
+ \bigg(\frac{256}{27}\bigg)  \Cf  \nf^2
+ \bigg(\frac{7744}{27}\bigg)  \Ca^2  \Cf
+ \bigg(\frac{17152}{9}
- 512  \z2
+ \bigg(
- \frac{14080}{9}\bigg)  \Lqf\bigg)  \Ca  \Cf^2
+ \bigg(\bigg(\frac{8704}{9}
+ \bigg(
- \frac{2048}{3}\bigg)  \Lqf\bigg)  \Cf^3\bigg)  \rgq  {\color{MidnightBlue} \bigg\} }      
+ \D2     {\color{MidnightBlue} \bigg\{ }  \bigg(
- \frac{28480}{27}
+ \frac{704}{3}  \z2
+ \bigg(\frac{3872}{9}\bigg)  \Lqf\bigg)  \Ca^2  \Cf
+ \bigg(
- \frac{1696}{27}
- \frac{2048}{3}  \z2
+ \bigg(
- \frac{2752}{9}\bigg)  \Lqf
+ \bigg(192\bigg)  \Lqf^2\bigg)  \Cf^2  \nf
+ \bigg(
- \frac{640}{27}
+ \bigg(\frac{128}{9}\bigg)  \Lqf\bigg)  \Cf  \nf^2
+ \bigg(\frac{4864}{27}
+ 1344  \z3
+ \frac{11264}{3}  \z2
+ \bigg(\frac{19744}{9}
- 768  \z2\bigg)  \Lqf
+ \bigg(-1056\bigg)   \Lqf^2\bigg)  \Ca  \Cf^2
+ \bigg(\frac{9248}{27}
- \frac{128}{3}  \z2
+ \bigg(
- \frac{1408}{9}\bigg)  \Lqf\bigg)  \Ca  \Cf  \nf
+ \bigg(10240  \z3
+ \bigg(
- \frac{15872}{3}
- 4608  \z2\bigg)  \Lqf
+ \bigg(256\bigg)  \Lqf^3
+ \bigg(2176\bigg)  \Lqf^2\bigg)   \Cf^3
+ \bigg(\bigg(
- \frac{11968}{27}
+ \bigg(\frac{2816}{9}\bigg)  \Lqf\bigg)  \Ca  \Cf^2
+ \bigg(\frac{2176}{27}
+ \bigg(
- \frac{512}{9}\bigg)  \Lqf\bigg)   \Cf^2  \nf
+ \bigg(\bigg(\frac{4352}{3}\bigg)  \Lqf
+ \bigg(-1024\bigg)  \Lqf^2\bigg)  \Cf^3\bigg)  \rgq  {\color{MidnightBlue} \bigg\} }      
+ \D1     {\color{MidnightBlue} \bigg\{ }  \bigg(
- \frac{604948}{81}
- 4160  \z3
- \frac{1664}{3}  \z2
+ \frac{3648}{5}  \z2^2
+ \bigg(
- \frac{704}{3}\bigg)   \Lqf^3
+ \bigg(
- \frac{400}{9}
- 256  \z2\bigg)  \Lqf^2
+ \bigg(\frac{37456}{9}
+ 512  \z3
+ \frac{8768}{3}  \z2\bigg)   \Lqf\bigg)  \Ca  \Cf^2
+ \bigg(
- \frac{32816}{81}
+ 384  \z2
+ \bigg(
- \frac{704}{9}\bigg)  \Lqf^2
+ \bigg(\frac{9248}{27}
- \frac{128}{3}  \z2\bigg)  \Lqf\bigg)   \Ca  \Cf  \nf
+ \bigg(\frac{1600}{81}
- \frac{256}{9}  \z2
+ \bigg(
- \frac{640}{27}\bigg)  \Lqf
+ \bigg(\frac{64}{9}\bigg)  \Lqf^2\bigg)  \Cf  \nf^2
+ \bigg(\frac{32768}{27}
+ 1280  \z3
- \frac{256}{9}  \z2
+ \bigg(
- \frac{7552}{9}
- \frac{1792}{3}  \z2\bigg)  \Lqf
+ \bigg(\frac{128}{3} \bigg)  \Lqf^3
+ \bigg(\frac{608}{9}\bigg)  \Lqf^2\bigg)  \Cf^2  \nf
+ \bigg(\frac{124024}{81}
- 704  \z3
- \frac{12032}{9}  \z2
+ \frac{704}{5}  \z2^2
+ \bigg(
- \frac{28480}{27}
+ \frac{704}{3}   \z2\bigg)  \Lqf
+ \bigg(\frac{1936}{9}\bigg)  \Lqf^2\bigg)  \Ca^2  \Cf
+ \bigg(\frac{172372}{27}
- 1984  \z3
+ \frac{42656}{9}  \z2
- \frac{14208}{5}  \z2^2
+ \bigg(
- \frac{137200}{27}
+ 11008  \z3
- \frac{5504}{3}  \z2\bigg)  \Lqf
+ \bigg(
- 736
- 2048  \z2\bigg)  \Lqf^2
+ \bigg(\frac{2176}{3}\bigg)  \Lqf^3 \bigg)  \Cf^3
+ \bigg(\bigg(
- \frac{287968}{81}
- \frac{12800}{9}  \z2
+ \bigg(
- \frac{3328}{9}\bigg)  \Lqf^2
+ \bigg(
- \frac{1024}{3}\bigg)  \Lqf^3
+ \bigg(\frac{95360}{27}
+ \frac{2048}{3}  \z2\bigg)  \Lqf\bigg)  \Cf^3
+ \bigg(
- \frac{16064}{27}
+ \frac{1024}{3}  \z2
+ \bigg(
- \frac{1024}{9}\bigg)  \Lqf^2
+ \bigg(\frac{12416}{27}\bigg)  \Lqf\bigg)  \Cf^2  \nf
+ \bigg(\frac{198112}{81}
- \frac{9728}{9}  \z2
+ \bigg(
- \frac{53120}{27}
+ \frac{512}{3}  \z2\bigg)  \Lqf
+ \bigg(\frac{1408}{3}\bigg)  \Lqf^2\bigg)  \Ca  \Cf^2\bigg)  \rgq
+ \bigg(\bigg(\frac{18496}{81}
+ \frac{2048}{3}  \z2
+ \bigg(
- \frac{8704}{27}\bigg)  \Lqf
+ \bigg(\frac{1024}{9}\bigg)  \Lqf^2\bigg)  \Cf^3\bigg)   \rgq^2  {\color{MidnightBlue} \bigg\} }      
+ \D0     {\color{MidnightBlue} \bigg\{ }  \bigg(
- \frac{594058}{729}
- 384  \z5
+ \frac{40144}{27}  \z3
+ \frac{98224}{81}  \z2
- \frac{352}{3}  \z2   \z3
- \frac{2992}{15}  \z2^2
+ \bigg(
- \frac{7120}{27}
+ \frac{176}{3}  \z2\bigg)  \Lqf^2
+ \bigg(\frac{968}{27}\bigg)  \Lqf^3
+ \bigg(\frac{62012}{81}
- 352  \z3
- \frac{6016}{9}  \z2
+ \frac{352}{5}  \z2^2\bigg)  \Lqf\bigg)  \Ca^2  \Cf
+ \bigg(
- \frac{24754}{243}
- \frac{5728}{9}  \z3
+ \frac{1760}{9}  \z2
- \frac{1472}{15}  \z2^2
+ \bigg(
- \frac{3784}{9}
- \frac{320}{3}  \z2\bigg)  \Lqf^2
+ \bigg(\frac{224}{3}\bigg)  \Lqf^3
+ \bigg(\frac{56768}{81}
+ 640  \z3
- \frac{1216}{9}  \z2\bigg)  \Lqf \bigg)  \Cf^2  \nf
+ \bigg(
- \frac{3712}{729}
+ \frac{320}{27}  \z3
+ \frac{640}{27}  \z2
+ \bigg(
- \frac{160}{27}\bigg)  \Lqf^2
+ \bigg(\frac{32}{27}\bigg)   \Lqf^3
+ \bigg(\frac{800}{81}
- \frac{128}{9}  \z2\bigg)  \Lqf\bigg)  \Cf  \nf^2
+ \bigg(\frac{125252}{729}
- \frac{2480}{9}  \z3
- \frac{29392}{81}  \z2
+ \frac{736}{15}  \z2^2
+ \bigg(
- \frac{16408}{81}
+ 192  \z2\bigg)  \Lqf
+ \bigg(
- \frac{352}{27}\bigg)  \Lqf^3
+ \bigg(\frac{2312}{27}
- \frac{32}{3}  \z2\bigg)  \Lqf^2\bigg)  \Ca  \Cf  \nf
+ \bigg(\frac{400768}{243}
+ \frac{20416}{9}  \z3
- \frac{10240}{9}  \z2
- 1472  \z2  \z3
+ \frac{1408}{3}  \z2^2
+ \bigg(
- \frac{357418}{81}
- \frac{4336}{3}  \z3
+ \frac{3488}{9}  \z2
+ \frac{1824}{5}  \z2^2\bigg)  \Lqf
+ \bigg(
- \frac{2992}{9}\bigg)   \Lqf^3
+ \bigg(\frac{19400}{9}
- 192  \z3
+ \frac{1216}{3}  \z2\bigg)  \Lqf^2\bigg)  \Ca  \Cf^2
+ \bigg(12288  \z5
- \frac{63488}{9}  \z3
- 6144  \z2  \z3
+ \bigg(
- \frac{68600}{27}
+ 2432  \z3
- \frac{2752}{3}   \z2\bigg)  \Lqf^2
+ \bigg(\frac{4624}{9}
- 256  \z2\bigg)  \Lqf^3
+ \bigg(\frac{86186}{27}
+ \frac{5728}{3}  \z3
+ \frac{21328}{9}  \z2
- \frac{7104}{5}  \z2^2\bigg)  \Lqf\bigg)  \Cf^3
+ \bigg(\bigg(
- \frac{109888}{243}
+ \frac{3808}{9}  \z3
+ \frac{11968}{27}  \z2
+ \bigg(
- \frac{7856}{9}
+ \frac{256}{3}  \z2\bigg)   \Lqf^2
+ \bigg(\frac{1408}{9}\bigg)  \Lqf^3
+ \bigg(\frac{124912}{81}
- \frac{896}{3}  \z3
- \frac{2560}{3}  \z2\bigg)  \Lqf\bigg)  \Ca   \Cf^2
+ \bigg(\frac{15232}{243}
- \frac{2176}{27}  \z2
+ \bigg(
- \frac{27680}{81}
+ \frac{2048}{9}  \z2\bigg)  \Lqf
+ \bigg(
- \frac{128}{3}\bigg)  \Lqf^3
+ \bigg(\frac{1888}{9}\bigg)  \Lqf^2\bigg)  \Cf^2  \nf
+ \bigg(\frac{17408}{9}  \z3
+ \bigg(
- \frac{143984}{81}
- \frac{4096}{3}  \z3
- \frac{6400}{9}  \z2\bigg)  \Lqf
+ \bigg(
- \frac{1280}{3}\bigg)  \Lqf^3
+ \bigg(\frac{47680}{27}
+ \frac{1024}{3}  \z2\bigg)  \Lqf^2\bigg)  \Cf^3\bigg)  \rgq
+ \bigg(\bigg(\bigg(
- \frac{4352}{27}\bigg)  \Lqf^2
+ \bigg(\frac{512}{9}\bigg)  \Lqf^3
+ \bigg(\frac{9248}{81}
+ \frac{1024}{3}  \z2\bigg)  \Lqf\bigg)   \Cf^3\bigg)  \rgq^2  {\color{MidnightBlue} \bigg\} } 
\end{autobreak}
\end{align}

\bibliographystyle{JHEP}
\bibliography{gr_dc_resum}
\end{document}